\documentclass[amsmath,amssymb,10pt,superscriptaddress,reprint, preprintnumbers, notitlepage,aps,prl,twocolumn]{revtex4-1}
\pdfoutput=1 
\usepackage[utf8]{inputenc}
\usepackage[english]{babel}
\usepackage{amsmath}
\usepackage{graphicx}
\usepackage{dcolumn}
\usepackage{pbox}
\usepackage{amssymb}
\usepackage{epsfig}
\usepackage[dvipsnames]{xcolor}
\usepackage{slashed}
\usepackage{amssymb}
\usepackage{ mathrsfs }
\usepackage{color}
\usepackage[font=small]{caption}
\usepackage[font=small]{subcaption}
\usepackage{url}
\definecolor{DarkBlue}{rgb}{0.7, 0.4, 1} 
\definecolor{Blue}{rgb}{0, 0.8, 0} 
\definecolor{MyLightBlue}{rgb}{0.5,0.7,1.9}
\definecolor{MyGreen}{rgb}{0.0,0.2, 0.0}
\definecolor{MyBrickRed}{rgb}{0, 0.5, 0.2}
\RequirePackage{hyperref}
\hypersetup{colorlinks, citecolor=Blue,linkcolor=DarkBlue, urlcolor=Green}

\newcommand{\bea}{\begin{eqnarray}}
\newcommand{\eea}{\end{eqnarray}}

\makeatletter

\makeatletter
\renewcommand\@makecaption[2]{%
  \par
  \vskip\abovecaptionskip
  \begingroup
  
   \small\rmfamily
    \begingroup
     \samepage
     \flushing
     \let\footnote\@footnotemark@gobble
     \@make@capt@title{#1}{#2}\par
    \endgroup
  \endgroup
  \vskip\belowcaptionskip
}
\makeatother

\DeclareUnicodeCharacter{2212}{-}
\begin{document}
\title{Testing neutrino mass hierarchy under type-II seesaw scenario in $U(1)_X$ from colliders}
\author{Arindam Das}
\email{arindamdas@oia.hokudai.ac.jp}
\affiliation{Institute for the Advancement of Higher Education, Hokkaido University, Sapporo 060-0817, Japan}
\affiliation{Department of Physics, Hokkaido University, Sapporo 060-0810, Japan}
\author{Puja Das}
\email{pdas1@crimson.ua.edu}
\affiliation{Department of Physics and Astronomy, University of Alabama, Tuscaloosa, AL35487, USA}
\author{Nobuchika Okada}
\email{okadan@ua.edu}
\affiliation{Department of Physics and Astronomy, University of Alabama, Tuscaloosa, AL35487, USA}

\begin{abstract}
The origin of tiny neutrino mass is a long standing unsolved puzzle of the Standard Model (SM), 
which allows us to consider scenarios beyond the Standard Model (BSM) in a variety of ways. 
One of them being a gauge extension of the SM that may be realized as in the form of an anomaly free, 
general $U(1)_X$ extension of the SM, where an $SU(2)_L$ triplet scalar with a $U(1)_X$ charge 
is introduced to have Dirac Yukawa couplings with the SM lepton doublets. 
Once the triplet scalar develops a Vacuum Expectation Value (VEV), light neutrinos acquire their tiny Majorana masses.  
Hence, the decay modes of the triplet scalar have direct connection to the neutrino oscillation data for different neutrino mass hierarchies. 
After the breaking of the $U(1)_X$ gauge symmetry, a neutral $U(1)_X$ gauge boson $(Z^\prime)$ acquires mass, 
which interacts differently with the left and right handed SM fermions. 
Satisfying the recent LHC bounds on the triplet scalar and $Z^\prime$ boson productions, we study the pair production of the triplet scalar
at LHC, 100 TeV proton proton collider FCC, $e^-e^+$ and $\mu^-\mu^+$ colliders followed by its decay into dominant dilepton modes 
whose flavor structure depend on the neutrino mass hierarchy. 
Generating the SM backgrounds, we study the possible signal significance of four lepton final states from the triplet scalar pair production. 
We also compare our results with the purely SM gauge mediated triplet scalar pair production followed by four lepton final states, 
which could be significant only in $\mu^- \mu^+$ collider.

\noindent 
\end{abstract}

\maketitle

\section{Introduction}
Observation of neutrino mass and flavor mixing \cite{ParticleDataGroup:2020ssz}  is one of the suitable aspects 
where Standard Model (SM) falls short of explaining it and hence beyond the SM (BSM) scenarios step in. 
Among a variety of simple but interesting aspects, neutrino mass can be explained extending the SM 
with an $SU(2)_L$ triplet scalar field with hyper-charge $Y=+2$, commonly known as type-II seesaw scenario \cite{Schechter:1981cv,Magg:1980ut,Cheng:1980qt,Lazarides:1980nt,Mohapatra:1980yp}. 
Apart from the particle extension of the SM, there is also a promising aspect where the SM can be extended with a $U(1)_{B-L}$ gauge group \cite{Pati:1973uk,Davidson:1978pm,Davidson:1979wr,Marshak:1979fm,Mohapatra:1980qe}. 
In the light of $U(1)$ extension of SM, we study a general $U(1)_X$ extension of the SM 
where an $SU(2)_L$ triplet scalar with a $U(1)_X$ charge is introduced to participate in the neutrino mass generation mechanism,
which has been proposed in Ref.~\cite{Okada:2022cby}. 
The anomalies due to the charges of the SM fermions under general $U(1)$ scenarios are cancelled 
by the introduction of three generations of SM-singlet right handed neutrinos (RHNs) 
which have flavor inhomogeneous charges under the general $U(1)_X$ gauge group preventing them to have Dirac Yukawa couplings
with the SM Higgs and lepton doublets. 
However, neutrino mass generation can be achieved through the $SU(2)_L$ triplet scalar field, which simply generates Majorana-type 
non-vanishing neutrino mass through the Yukawa couplings with the SM lepton doublets, not requiring any Dirac-type mass. 
Hence the smallness of neutrino mass is achievable through the Vacuum Expectation Value (VEV) of the triplet scalar and the corresponding Yukawa couplings. 
Note that the RHNs in this case do not participate in the neutrino mass generation mechanism, rather participate in Dark Matter (DM) phenomenology \cite{Okada:2022cby}. 
An additional interesting feature of a general $U(1)_X$ extension is the presence of a neutral BSM gauge boson $Z^\prime$ 
which has different interactions with the left and right handed SM fermions manifesting chiral nature of the model.

Addressing  the above facts, we propose a scenario that not only provides a correct degree of neutrino masses and flavor mixings
but is also a testable scenario at the Large Handron Collider (LHC) and other future colliders. 
In our model set-up, the triplet scalar acquires a VEV through a trilinear coupling with a BSM $SU(2)_L$ doublet scalar field
having a $U(1)_X$ charge which forbids its coupling to the SM fermions. 
The smallness of the VEV governs tiny neutrino mass satisfying the neutrino oscillation data. 
It should be noted that the smallness of neutrino masses also ensures a small VEV of the BSM doublet scalar. 
The $U(1)_X$ symmetry is spontaneously broken by a VEV of an SM-singlet scalar field introduced with a unit $U(1)_X$ charge.
It further generates mass of the pseudo-scalar and the Dirac mass term of the RHNs.  
One of the RHNs is massless at the renormalization level triggering the fact that it could be considered as a Dark Radiation (DR) 
in the universe, and the remaining massive modes could compose a viable DM candidate \cite{Okada:2022cby}. 
DRs could fix the tension between the Hubble parameters \cite{DiValentino:2021izs} obtained from PLANK \cite{Planck:2018vyg} and SH0ES \cite{delaMacorra:2021hoh} collaborations.

This economical extension of the SM explains the aspect of neutrino mass generation mechanism
which could be tested directly at the high energy colliders where the doubly charged component in the $SU(2)_L$ triplet scalar
is produced from the resonance of $Z^\prime$ boson followed by same-sign dilepton final states. 
Here, note that the BSM gauge boson $Z^\prime$ plays a crucial role to search for the BSM scalars. 
Recent ATALS experiment have ruled out the possibility of a doubly charged scalar below $1$ TeV at 139 fb$^{-1}$ \cite{ATLAS:2022pbd} from multilepton final states at $\sqrt{s}=13$ TeV. 
As a result a four lepton final state is challenging to observe at the colliders if doubly charged scalars are produced
through the SM gauge interactions. 
Therefore, we emphasize on the doubly charged scalar production through a BSM propagator $Z^\prime$. 
However, the recent dilepton searches from a narrow $Z^\prime$ boson resonance at the LHC \cite{ATLAS:2019erb, CMS:2021ctt} 
provide us with a severe limit on (sequential SM) $Z^\prime$ boson mass, $M_{Z^\prime} > 5.2$ TeV, 
with the integrated luminosity of around 140 fb$^{-1}$. 
Thus, $Z^\prime$ mediated doubly charged scalar production is also challenging, however, if the dilepton bound is relaxed
with general $U(1)_X$ couplings, then a heavy doubly charged scalar can be produced from the heavy $Z^\prime$ boson 
dominating over the processes mediated by SM gauge bosons.
This is a noticeable feature of our scenario. 
Following the constrains on the doubly charged scalars, the $\rho$-parameter bound, $\rho= 1.00038 \pm 0.00020$, 
providing a limit on triplet VEV, $v_{\Delta} \leq 0.78(2.6)$ GeV at 2(3)$\sigma$ \cite{ParticleDataGroup:2020ssz}, 
and non-observation of exotic decay of the SM $Z$ boson ruling out a doubly charged scalar lighter than half of  the $Z$ boson mass \cite{Kanemura:2013vxa}, we study this scenario to test neutrino mass hierarchy and lepton flavor violating (LFV) processes considering BR($\mu\to 3e$)$< 1.0 \times 10^{-12}$ from SINDRUM \cite{SINDRUM:1987nra} and BR($\mu\to e\gamma$) $< 4.2\times 10^{-12}$ from  MEG \cite{MEG:2016leq} experiments, respectively. These LFV processes are proportional to roughly fourth power of the Dirac Yukawa coupling between triplet scalar and lepton doublet. As a result it could provide a lower limit on $v_\Delta$ for doubly charged scalar mass around $\mathcal{O}(1)$ TeV and its decay modes \cite{Mandal:2022zmy,Mandal:2022ysp,Das:2023tna}.

Based on the type-II seesaw scenario in the general $U(1)_X$ extension of the SM along with the recent experimental limits, 
our study focuses on the production of the doubly charged triplet scalar from $Z^\prime$ boson in the context of LHC, 100 TeV proton proton collider FCC,  $e^- e^+$ and $\mu^- \mu^+$ colliders followed by the decay of doubly charged triplet scalar into same-sign dilepton 
final states; though its Yukawa coupling is directly related to the neutrino oscillation data with normal and inverted hierarchies. 
We also consider the doubly charged scalar pair production through the SM gauge interactions to compare with the $Z^\prime$ mediated process. 
Studying generic SM backgrounds for different colliders, we compare signal and background events 
to investigate the role of triplet scalar in generating tiny neutrino mass.

\section{Model}
We work on the SM$\otimes U(1)_X$ framework \cite{Okada:2022cby},
where the SM quark fields transform as $q_L^i=\{3,2,\frac{1}{6}, x_q=\frac{1}{6} x_H+\frac{1}{3}\}$, $u_R=\{3,1,\frac{2}{3}, x_u=\frac{2}{3} x_H+\frac{1}{3}\}$, $d_R^i=\{3,1,-\frac{1}{3}, x_d=-\frac{1}{3}x_H+\frac{1}{3}\}$, respectively, 
whereas SM lepton fields transforms as $\ell_L^i=\{1,2,-\frac{1}{2}, x_\ell=-\frac{1}{2}x_H-1\}$, $e_R^i=\{1,1,-1, x_e=-x_H-1\}$, respectively, where $i=1, 2, 3$ represents three generations. 
In this framework, the SM Higgs field transforms as $H=\{1,2,\frac{1}{2}, x_h=\frac{1}{2}x_H\}$. 
We introduce three SM-singlet RHNs to cancel gauge and mixed gauge-gravity anomalies which transform as $N_R^{\{j=1,2\}}=\{1,1,0,x_\nu^{\{j=1,2\}}=-4\}$, $N_R^3=\{1,1,0, x_\nu^3=5\}$ with inhomogeneous $U(1)_X$ charges, 
and an SM-singlet $U(1)_X$ scalar which transforms as $\Phi=\{1,1,0, 1\}$ to break the $U(1)_X$ symmetry. 
Finally, the additional scalar fields transform as $\tilde{H}=\{1,2,\frac{1}{2}, \frac{1}{2}x_H+1\}$ and $\Delta=\{1,3,1, x_{\Delta}=x_H+2\}$, respectively. 
The $U(1)_X$ charge of the fields can be defined as the linear combination of $U(1)_{B-L}$ and $U(1)_Y$, 
so that the model is free from the $U(1)_X$ related anomalies since $U(1)_{B-L}$ and $U(1)_Y$ are separately anomaly-free. 
Due to this charge assignment and $U(1)_X$ symmetry, RHNs cannot have Dirac Yukawa coupling with $\ell_L$ and $H$, 
and hence in this model setup, light neutrino mass cannot be generated via the type-I seesaw mechanism.  
We economically introduce $\Delta$ to have Yukawa couplings with $\ell_L$, which generates Majorana-type light neutrino mass 
through its VEV. 
This is commonly known as the type-II seesaw mechanism.
Although the other scalar doublet $\tilde{H}$ does not have any Yukawa coupling with the SM leptons and quarks like the type-I 2HDM \cite{Branco:2011iw}, its coupling with $\Delta$ is crucial to generate a small VEV for $\Delta$ \cite{Okada:2022cby}.

The Yukawa interactions for the triplet and singlet scalar fields from the relevant part of the Lagrangian are given as 
\bea
\mathcal{L}^Y=-\frac{1}{\sqrt{2}} Y^{ij} \overline{{\ell_L^i}^C}.\Delta \ell_L^j-\sum_{i=1,2} \tilde{Y}^{i} \Phi \overline{{N_R^i}^C} N_R^3+ h.c.,  
\label{Yuk1}
\eea
where $C$ denotes the charge conjugate, dot represents antisymmetric product of $SU(2)$ gauge group,
$Y$ and $\tilde{Y}$ are the Yukawa couplings, respectively. 
After a VEV of $\Delta$ is generated, Majorana-type left handed neutrino masses are generated as $\mathcal{M}_{\nu}=Y^{ij} v_{\Delta}$,  where $v_\Delta$ is the triplet VEV. 
After the breaking of $U(1)_X$ symmetry through the VEV of $\Phi$, Dirac masses of the RHNs can be generated. 
We can see the one of the three mass eigenstates is massless and can be considered as DR, 
while the remaining two mass eigenstates form Dirac spinors, and the lighter eigenstate plays the role of DM in our universe \cite{Okada:2022cby}. 
As mentioned above, the $U(1)_X$ gauge symmetry forbids the type-I seesaw mechanism, and the neutrino mass 
can only be generated through type-II seesaw mechanism in our framework.

Due to general $U(1)_X$ gauge symmetry, the left and right handed fermions interact differently with the $Z^\prime$ (for $x_H \neq 0$), 
manifesting the chiral nature of the model. 
The corresponding Lagrangian can be written as 
\bea
\mathcal{L}= g_X \sum_{i} Q^i_{L,R}\overline{f^i_{L,R}} \gamma_\mu {Z^\prime}^\mu f^i_{L,R}, 
\label{Yuk2}
\eea
where $i$ corresponds to the sum over quark and lepton states and their three generations, 
$Q_{L,R}$ are the charges for the left and right handed fermions under $U(1)_X$, and $g_X$ is the $U(1)_X$ gauge coupling.

Now we discuss the scalar sector of the model. 
The potential is given by Eq.~(\ref{eq.pot}). 
The $U(1)_X$ symmetry is spontaneously broken at a high scale via $\langle \Phi \rangle=\frac{v_\Phi}{\sqrt{2}}=\frac{m_{\Phi}^2}{\lambda_{\Phi}}$, where $v_\Phi \gg 246$ GeV. 
Hence $U(1)_X$ gauge boson $Z^\prime$ and physical state from the SM-singlet scalar acquire mass as $M_{Z^\prime}=g_X^2 v_\Phi^2$ and $\tilde{m}_{\Phi}^2= \lambda_{\Phi}v_{\Phi}^2$, respectively. 
Substituting $\langle \Phi \rangle>=\frac{v_\Phi}{\sqrt{2}}$ into Eq.~(\ref{eq.pot}) and rearranging terms, 
we obtain the low energy effective potential below $v_\Phi$ which is given by Eq.~(\ref{eq1.pot}). 
From the last term in the low energy effective potential, we find $-\tilde{m}^2_{H\tilde{H}}= \lambda \frac{\Phi}{\sqrt{2}}$. 
In contrast to the 2HDM, $(\tilde{H}^\dagger H)^2$ term is absent due to the $U(1)_X$ symmetry, 
however, the term containing $m_{H\tilde{H}}^2$ is essential for generating the CP-odd scalar boson masses and
removing dangerous Nambu-Goldstone bosons from the mass spectrum. 
Using the stationary conditions on Eq.~(\ref{eq1.pot}), we can express $\tilde{m}_{i}^2$ with $i=H, \tilde{H}, \Delta$ in terms of the model parameters given by Eq.~(\ref{eq.st}). 
From Eq.~(\ref{eq.type-II}), we can see that if the trilinear coupling between $\tilde{H}$ and $\Delta$ and the VEV of the BSM scalar doublet are very small, type-II seesaw mechanism naturally occurs through a tiny $v_\Delta$, which 
easily satisfies the $\rho$ parameter constraint. 
Such a small $v_\Delta$ enhances the same-sign dilepton branching modes from $\Delta^{\pm \pm}$. The mass eigenvalues of neutral, charged scalars and pseudoscalars are given in Eq.~(\ref{eq.mass}).

\section{Results and Discussions} 
\begin{figure*}
\centering
\includegraphics[scale=0.37]{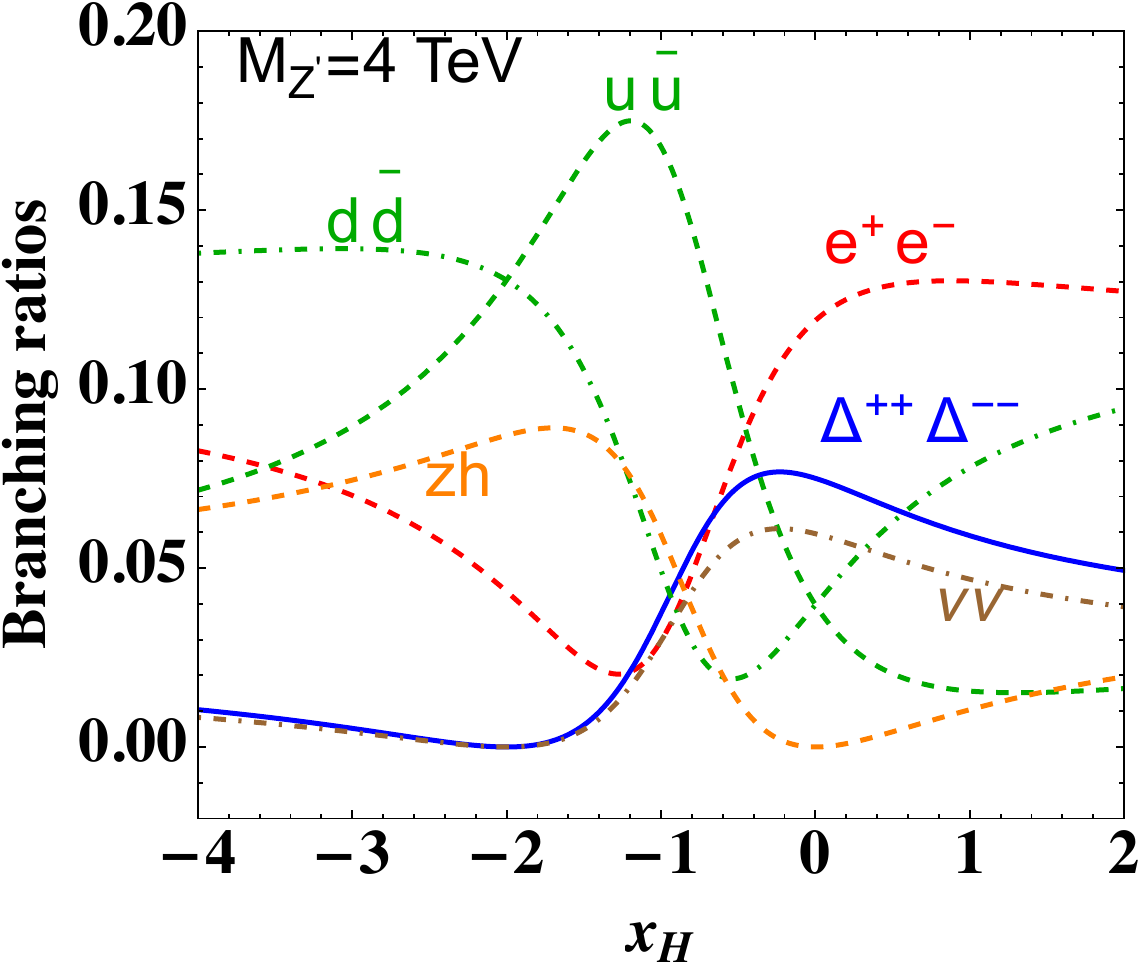}
\includegraphics[scale=0.36]{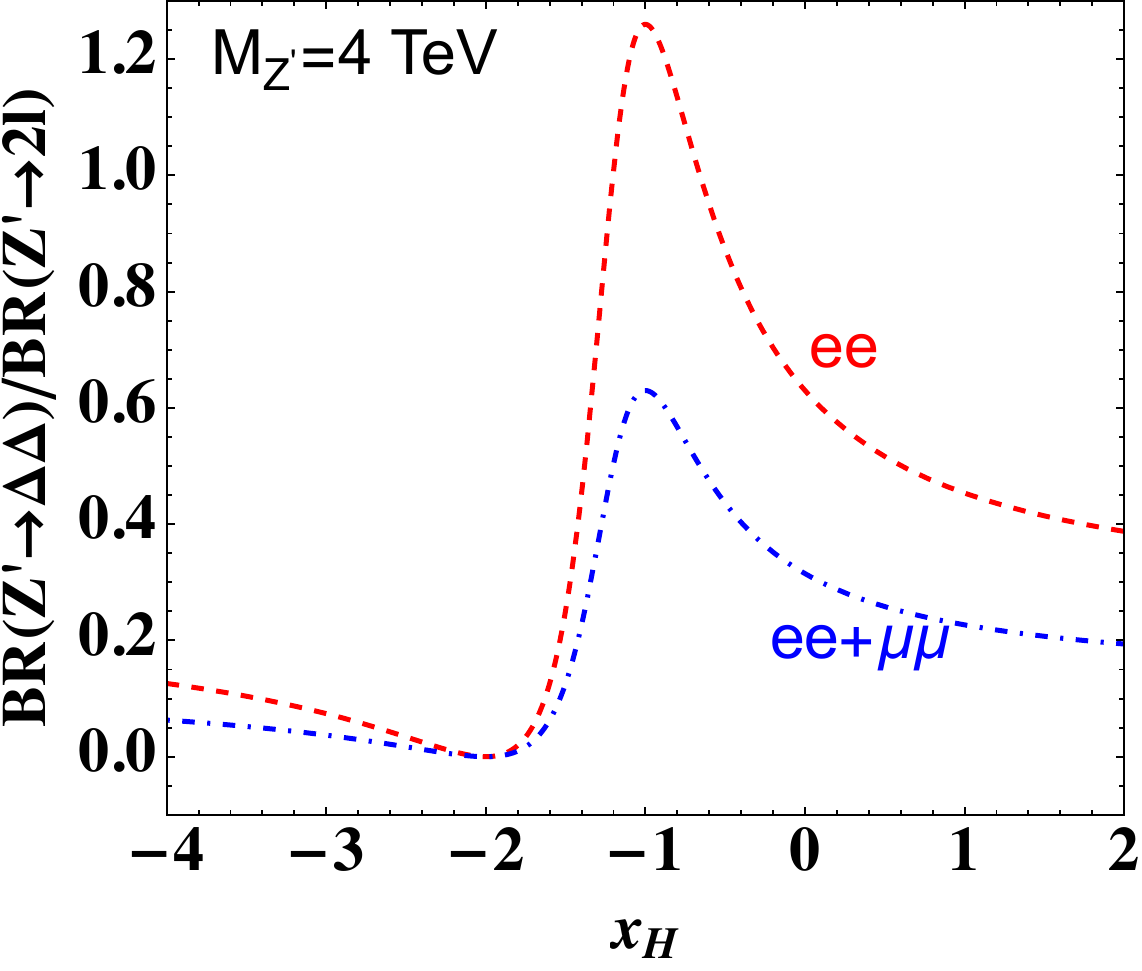}
\includegraphics[scale=0.37]{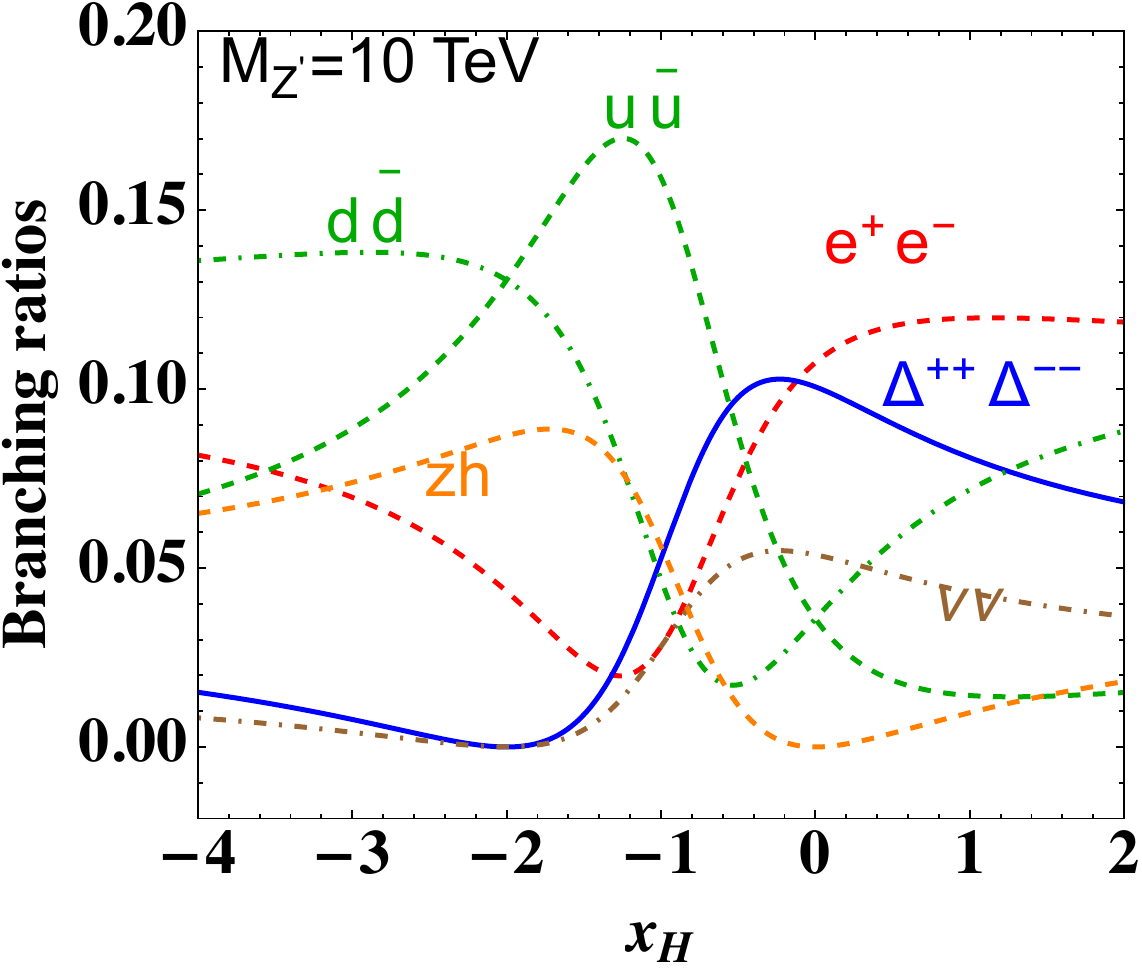}
\includegraphics[scale=0.36]{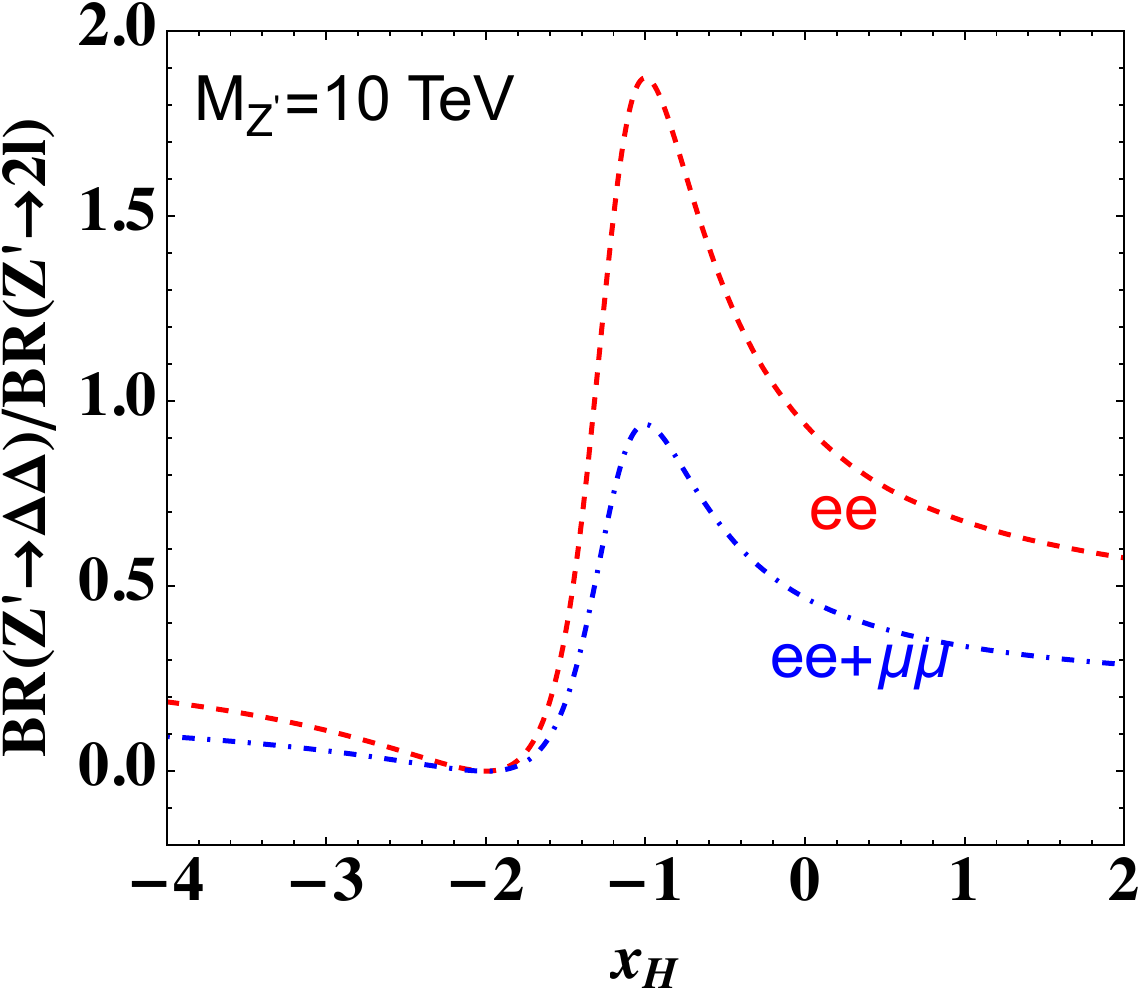}
\caption{
Branching ratios of $Z^\prime$ into different modes (left panel) and the ratio of branching ratios of the triplet 
to the electron positron (red) and electron positron plus muons (right panel) for different $M_{Z^\prime}$. 
We have used $m_{\Delta^{\pm \pm}}=1.03$ TeV. }
\label{fig:limZp2}
\end{figure*}

We show the branching ratios of $Z^\prime$ boson into different modes in the left panel and the ratio of branching ratios of $Z^\prime \to \Delta^{++} \Delta^{--}$ to $e^-e^+$ and $(e^+e^-+\mu^-\mu^+)$ in the right panel of Fig.~\ref{fig:limZp2}, respectively, for $m_{\Delta^{\pm \pm}}=1.03$ and $M_{Z^\prime}=4(10)$ TeV in the upper (lower) panel 
(we use $m_{\Delta^{\pm \pm}}=1.03$ TeV for our analysis through out of this paper).
From the left panel we find that at $x_H=-1$, BR$(Z^\prime \to \Delta^{++} \Delta^{--})\simeq 4.1\%$ 
which is slightly higher than BR$(Z^\prime \to e^- e^+,~\nu\nu,~zh)$ mode and comparable with BR$(Z^\prime \to d \bar{d})$, 
whereas BR$(Z^\prime \to u \bar{u})$ dominates over these branching ratios. 
In case of $x_H=0$, which is commonly known as the B$-$L scenario, 
BR$(Z^\prime \to \Delta^{++} \Delta^{--})\simeq 7.5\%$ is higher than all other modes except $e^- e^+$ case. 
For $x_H=1$, BR$(Z^\prime \to \Delta^{++} \Delta^{--})\simeq 6.0\%$ and it dominates over all modes 
except $e^- e^+$ and $d \bar{d}$ modes. 
Subsequently from the right panel, we find that BR$(Z^\prime \to \Delta^{++} \Delta^{--})$ is 1.3 times BR$(Z^\prime \to e^-e^+)$ 
and 0.65 times BR$(Z^\prime \to e^-e^+ +\mu^- \mu^+)$ at $x_H=-1$, 
whereas for  $x_H=0$ these are 0.63 and 0.32, respectively and for $x_H=1$ these are 0.46 and 0.23, respectively.

In case of $M_{Z^\prime}=10$ TeV, we find that BR$(Z^\prime \to \Delta^{++} \Delta^{--})=10\%$ for $x_H=0$ which dominates over all the modes and almost comparable to $Z^\prime \to e^+ e^-$ mode.
Also for $x_H=1$, BR$(Z^\prime \to \Delta^{++} \Delta^{--})=8\%$ which dominates over all the decay modes of $Z^\prime$ except $e^+ e^-$, 
whereas that for $x_H=-1$ becomes $5.3\%$ which is comparable to the $Zh$ mode, however, dominated by the BR$(Z^\prime \to u \bar{u})$ decay mode. Subsequently in the lower right panel, we find that BR$(Z^\prime \to \Delta^{++} \Delta^{--})$ is roughly two times more than  BR$(Z^\prime \to e^+ e^-)$, almost comparable to the  $Z^\prime \to e^+ e^-$ mode and for $x_H=1$ we find that BR$(Z^\prime \to \Delta^{++} \Delta^{--})$ is 0.67 times BR$(Z^\prime \to e^+ e^-)$, respectively. 
When we compare BR$(Z^\prime \to \Delta^{++} \Delta^{--})$ to BR$(Z^\prime \to e^-e^+ +\mu^- \mu^+)$, the ratio becomes half of BR$(Z^\prime \to e^-e^+)$ for $x_H=-1$, 0 and 1. 
We observe such changes due to the fact that the factor $\frac{m_{\Delta^{++}}}{M_{Z^\prime}}$ decreases with the increase in $M_{Z^\prime}$ for fixed $m_{\Delta^{++}}$. Therefore branching ratio of $Z^\prime \to \Delta^{++} \Delta^{--}$ increases in case of $M_{Z^\prime}=10$ TeV from 4 TeV. For these $U(1)_X$ charges, we study the doubly charged scalar production from $Z^\prime$ mediated processes, followed by its decay into same-sign dilepton modes. Note that at $x_H=-2$, BR$(Z^\prime \to \Delta^{++} \Delta^{--})$ becomes zero because the $U(1)_X$ charge of the triplet scalar vanishes at $x_H=-2$.

\begin{figure*}
\includegraphics[scale=0.23]{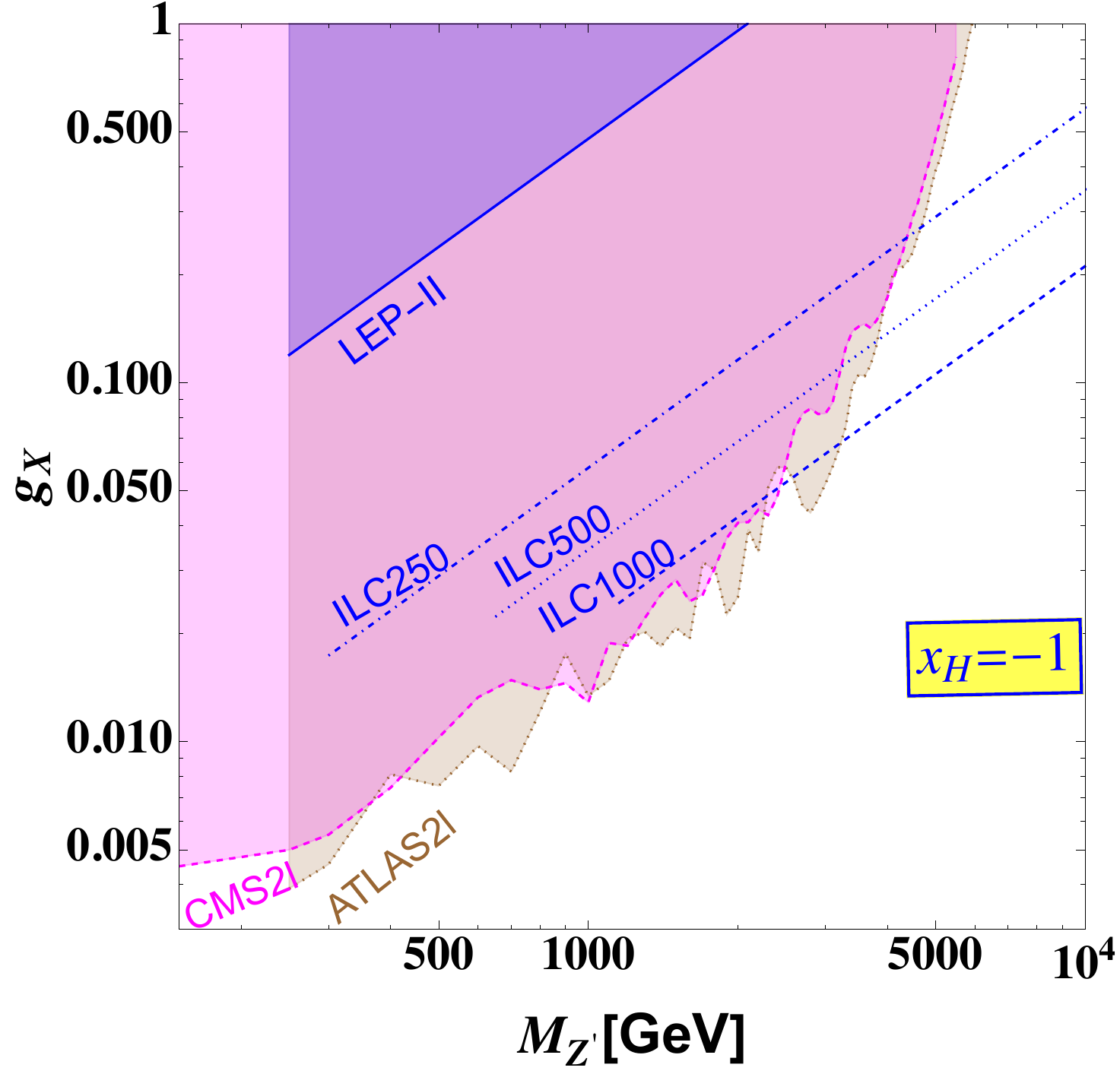}
\includegraphics[scale=0.23]{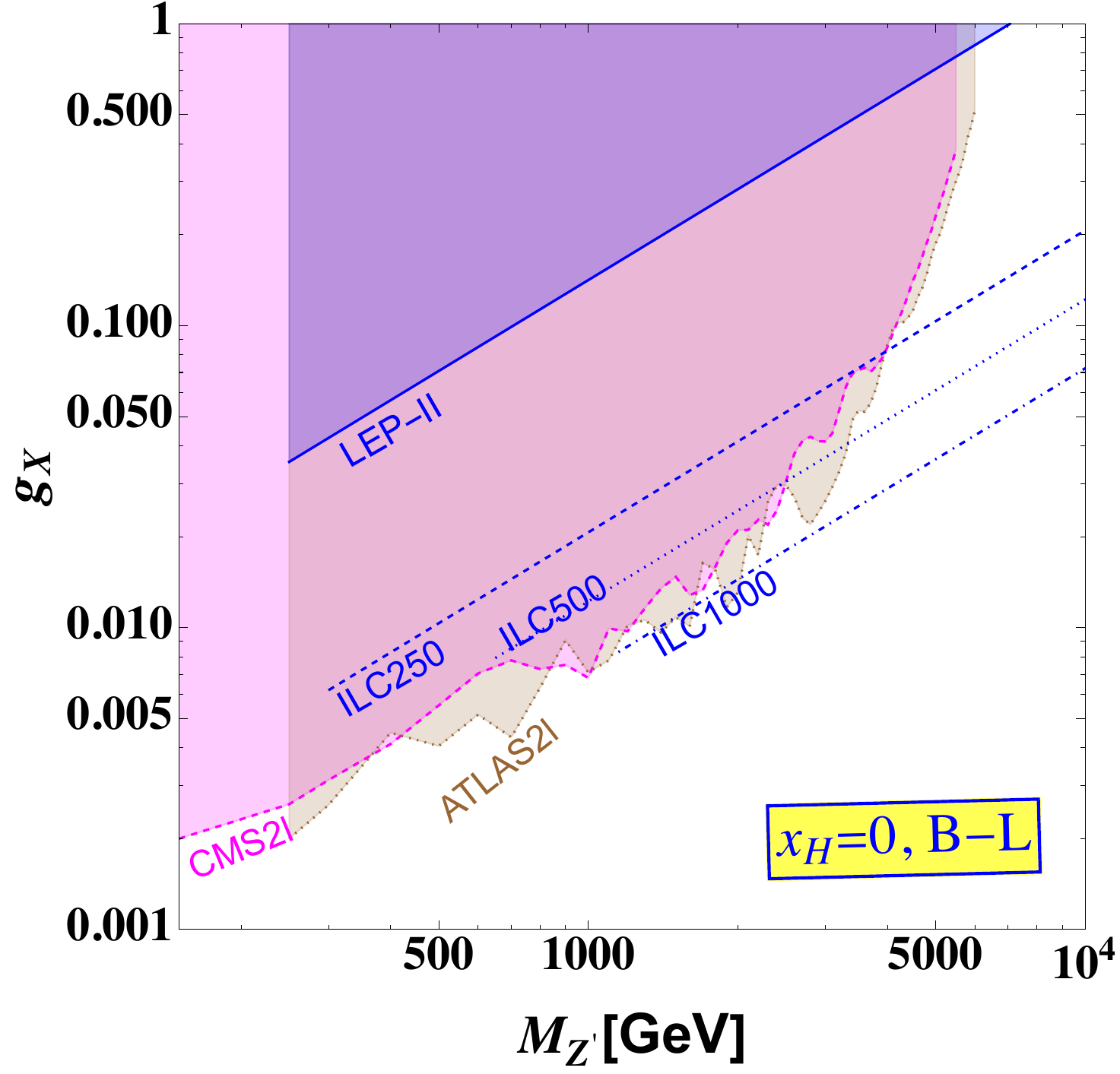}
\includegraphics[scale=0.23]{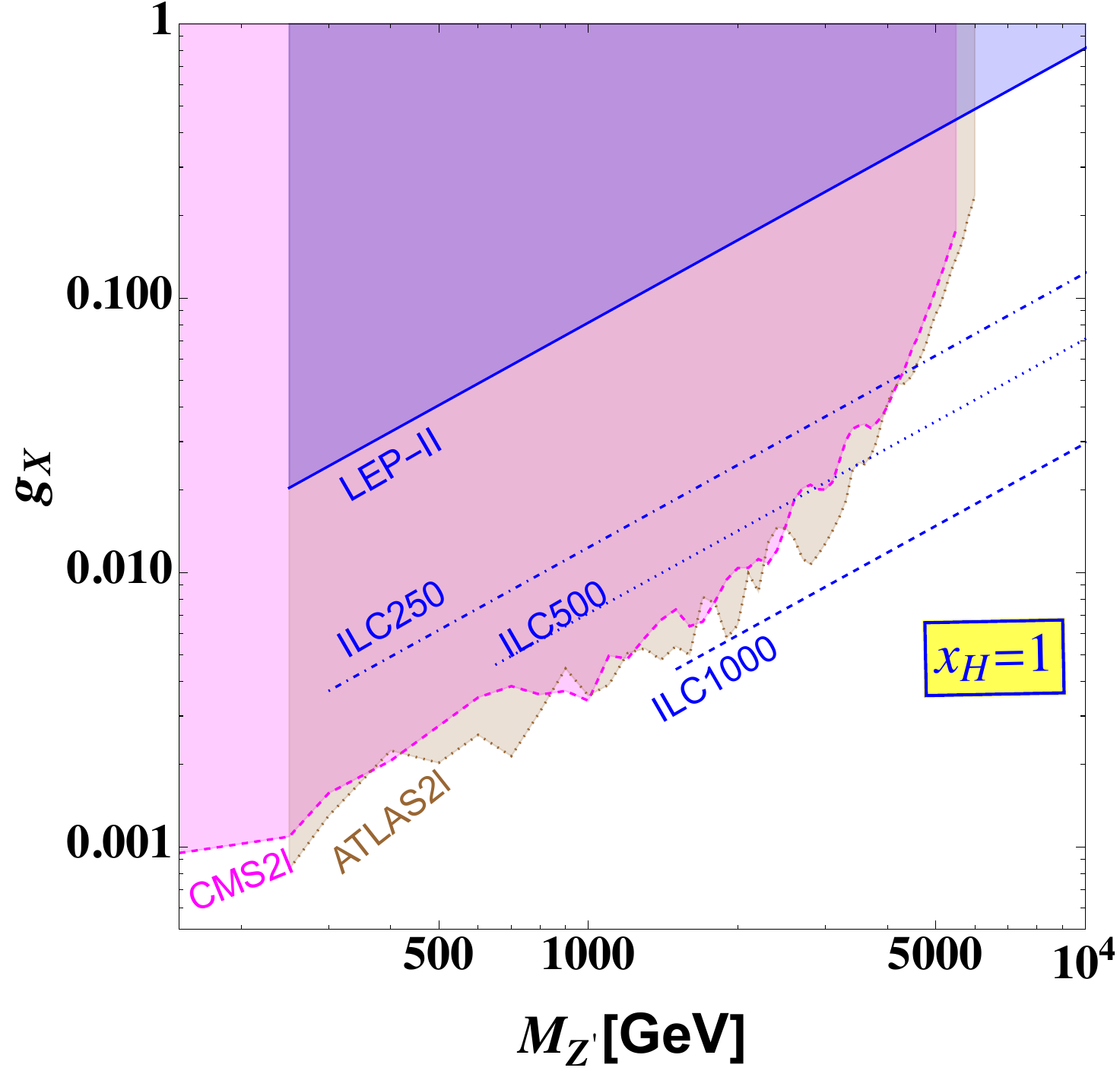}
\caption{Limits on $M_{Z^\prime}-g_X$ plane from LHC and $e^-e^+$ colliders for $x_H=-1$, $0$ (B$-$L), $1$, respectively.}
\label{fig:limZp}
\end{figure*}

Next we consider constraints on $M_{Z^\prime}-g_X$ plane for $x_H=-1, 0$ and $1$ by comparing the dilepton $(\ell= e,~\mu)$  
production cross section $(\sigma^\prime)$ in our model with those $(\sigma^{\rm{ATLAS, CMS}})$ from LHC \cite{ATLAS:2019erb, CMS:2021ctt} by using $g_X=g^{\prime}\sqrt{\frac{\sigma^{\rm{ATLAS/CMS}}}{\sigma^{\ell \ell}_{\rm LHC}}}$, 
where $g^\prime$ is the trial value of $U(1)_X$ gauge coupling to calculate $\sigma^{\ell \ell}_{\rm LHC}$ 
in the narrow width approximation (NWA) as
\bea
\sigma_{\rm LHC}^{\ell \ell}&=&3.89\times 10^{8} \times\frac{2 \pi^2}{3}  \int_{\frac{M_{Z^\prime}^2}{ E_{\rm LHC}^2}}^1 \frac{dx}{x E_{\rm LHC}^2} \nonumber \\
&&
\sum_{q, \overline{q}} \Big[ f_q (x, M_{Z^\prime}) f_{\overline{q}}(\frac{M_{Z^\prime}^2}{x E_{\rm LHC}^2}, M_{Z^\prime})\Big] 
\nonumber\\
&&
\frac{\Gamma_{Z^{\prime}(Z^{\prime}\to q\overline{q})}}{M_{Z^{\prime}}}\delta(\hat{s}-M_{Z^\prime}^2) f(x_H), 
\eea
where $f(x_H)=\Big(\frac{8+12 x_H+5x_H^2}{13+16x_H+10 x_H^2}\Big)$ and $f_{q(\bar{q})}$ being parton distribution function from CTEQ6L \cite{Pumplin:2002vw} for (anti)quark with a suitable $k-$factor of $0.947$ to match ATLAS \cite{ATLAS:2019erb} result which is slightly stronger than CMS \cite{CMS:2021ctt} prediction. The $x_H$ dependence is evolved from the interactions of SM quarks and leptons with $Z^\prime$ manifesting the chiral nature of the model. It further influences the partial decay widths of $Z^\prime$ given in Eq.~(\ref{pdw}). Estimated constrains are shown in Fig.~\ref{fig:limZp} for ATLAS and CMS results.

Limits on $g_X$ from LEP-II data using $M_{Z^\prime} \gg \sqrt{s}$  can be calculated for the B$-$L case considering the contact interaction $e^-e^+ \to f \bar{f}$ as
\bea
{\cal L}_{\rm eff}&=&\frac{g_X^2}{(1+\delta_{ef}) (\Lambda_{lm}^{f\pm})^2} 
\sum_{l,m=L,R} C_{lm}(\overline{e} \gamma^\mu P_l e)\nonumber \\
&&(\overline{f} \gamma_\mu P_m f) , 
\label{eq1}
\eea
where $g_X^2/4\pi$ is taken to be unity using the convention $\delta_{ef}=1\ (0)$ for $f=e$ ($f\neq e$), $C_{lm}=\pm 1$ or 0, and $\Lambda_{lm}^{f\pm}$ is the scale of contact interaction where constructive and destructive interference with SM processes $e^+e^-\to f\bar{f}$ \cite{Kroha:1991mn,Carena:2004xs} are represented by plus and minus signs in this account, respectively. 
Now we evaluate the $Z^\prime$ mediated matrix element of our $U(1)_X$ scenario as
\bea
&&\frac{g_X^2}{{M_{Z^\prime}}^2-s} [\overline{e} \gamma^\mu (\tilde{x}_\ell P_L+ \tilde{x}_e P_R) e] [\overline{f} \gamma_\mu (\tilde{x}_{f_L} P_L+ \nonumber \\  
&&\tilde{x}_{f_R} P_R) f] ,
\label{eq2}
\eea
where $\tilde{x}_{f_L}$ and $\tilde{x}_{f_R}$ are the corresponding $U(1)_X$ charges of the left handed $(f_L)$ and right handed $(f_R)$ fermions, respectively. 
By matching Eqs.~\eqref{eq1} and \eqref{eq2} we obtain
\bea
M_{Z^\prime}^2  \ \gtrsim \ \frac{{g_X}^2}{4\pi} |{x_{e_l}} x_{f_m}| (\Lambda_{lm}^{f\pm})^2
\label{Lim}
\eea
for $M_{Z^\prime}^2 \gg s$, where $\sqrt{s}=209$ GeV for LEP-II. 
We then estimate bounds on $M_{Z^\prime}/g_X$ from the LEP result 
by using different values of  $C_{lm}^{f\pm}$ for the general $U(1)_X$ scenario. 
Here we employ $95\%$ C.L. bounds on $\Lambda_{lm}^{f\pm}$ from \cite{ALEPH:2013dgf} for leptons and quarks 
with $lm=LL,\ RR, \ LR, \ RL, \ VV$ and $AA$, assuming the flavor universality on the contact interactions. 
Similarly, we estimate prospective limits on $M_{Z^\prime}/g_X$ for the B$-$L case at ILC with $\sqrt{s}=250$ GeV, $500$ GeV and $1$ TeV using the bounds on $\Lambda_{lm}^{f\pm}$ from \cite{LCCPhysicsWorkingGroup:2019fvj}. 
We find the bounds on $M_{Z^\prime}/g_X$ for $x_H=-1$ as 2.2 TeV from LEP-II and corresponding prospective bounds for ILC are 16.3 TeV, 26.3 TeV and 47.7 TeV for $\sqrt{s}=250$ GeV, 500 GeV and 1 TeV, respectively. 
For $x_H=0$ case, we find the bounds on $M_{Z^\prime}/g_X$ as 7.0 TeV for LEP-II while the ILC prospective bounds are 48.2 TeV, 81.6 TeV and 137.2 TeV for $\sqrt{s}=250$ GeV, 500 GeV and 1 TeV, respectively. 
The bounds on $M_{Z^\prime}/g_X$ for $x_H=1$ are found to be 11.1 TeV for LEP-II and prospective bounds are 79.0 TeV, 139.1 TeV and 238.2 TeV from ILC for $\sqrt{s}=250$ GeV, 500 GeV and 1 TeV, respectively. 
Corresponding limits on $M_{Z^\prime}-g_X$ plane are shown in Fig.~\ref{fig:limZp} at $95\%$ C.L.

\begin{figure*}
\centering
\includegraphics[scale=0.41]{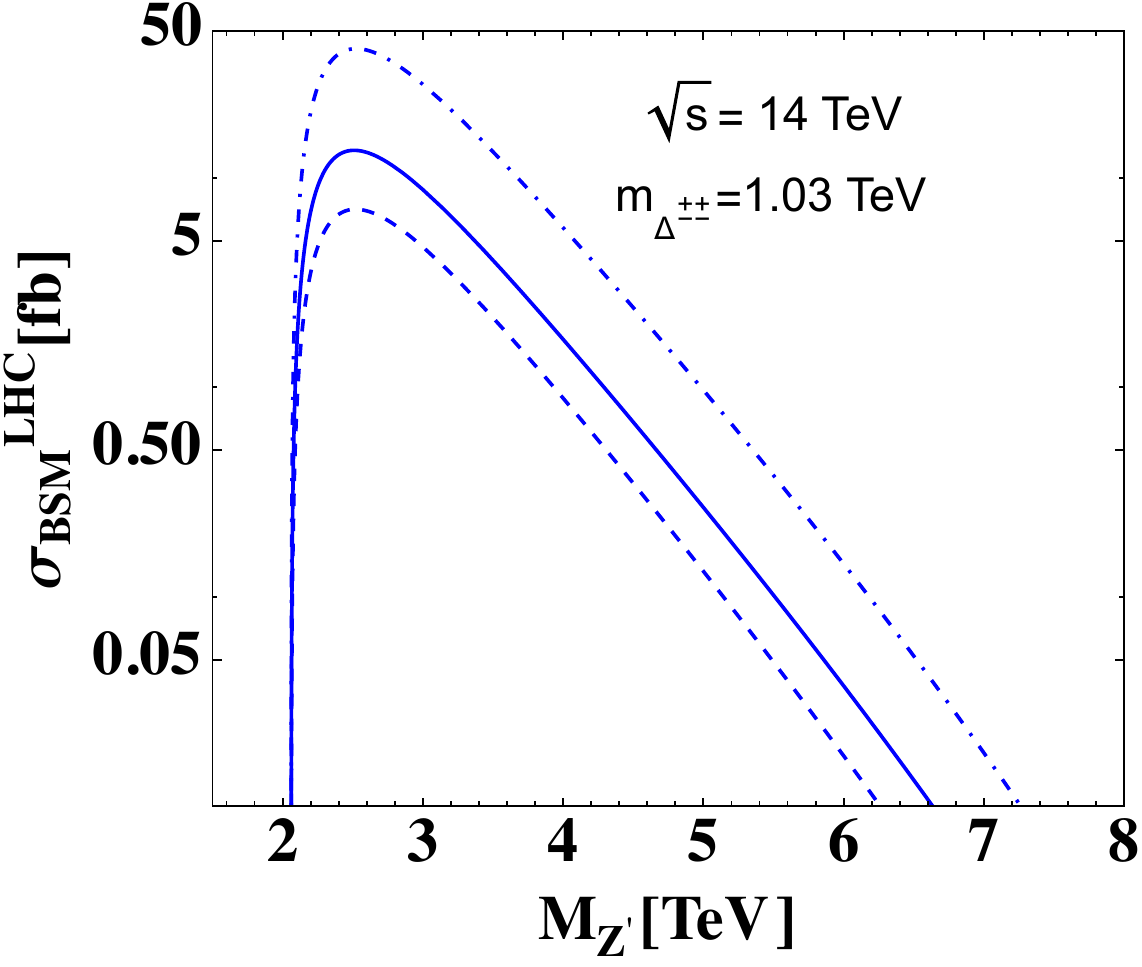}
\includegraphics[scale=0.42]{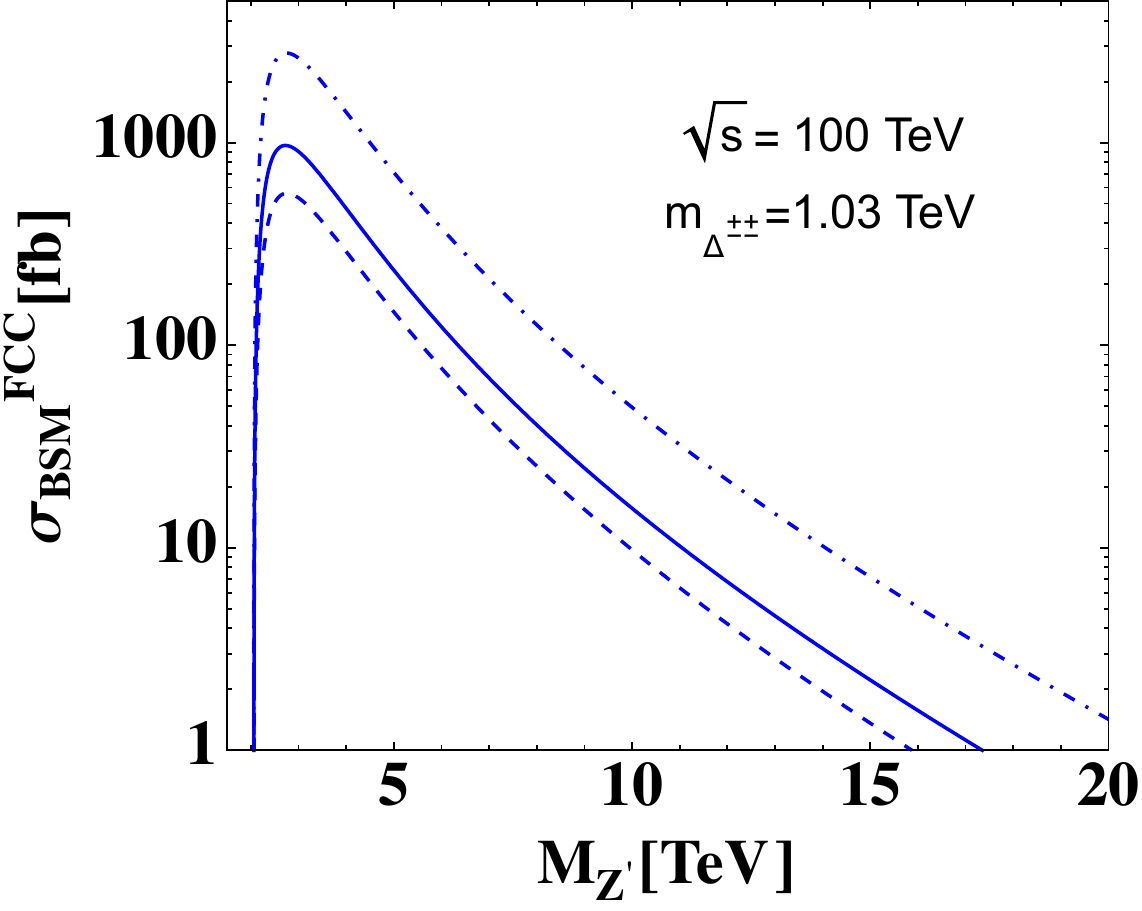}
\caption{$\Delta^{\pm\pm}$ production under general $U(1)_X$ scenario at the LHC and 100 TeV proton proton future circular collider (FCC) normalized by $g_X^2$ for $x_H=-1$ (dashed), $0$ (B$-$L) (solid) and $1$ (dot dashed) respectively.}
\label{fig:Xsec-1}
\end{figure*}

Now we consider the production of $\Delta^{\pm \pm}$ at LHC 
by considering the process $pp \to \Delta^{++} \Delta^{--}$ mediated by $Z^\prime$ boson
using NWA, the production cross section being normalized by $g_X^2$ can be written as
\bea
\sigma_{\rm BSM}^{\rm LHC}&=&3.89\times 10^{8}\times \frac{8 \pi^2}{3 } \int_{\frac{M_{Z^\prime}^2}{ E_{\rm LHC}^2}}^1 dx  \frac{1}{x {E_{\rm{LHC}}^2}}   \sum_{q, \bar{q}} \nonumber \\
&&\Big[f_q (x, M_{Z^\prime})f_{\overline{q}}
(\frac{M_{Z^\prime}^2}{x E_{\rm LHC}^2},M_{Z^\prime})\Big]\frac{\Gamma(Z^\prime\to q\overline{q})}{g_X^2 M_{Z^{\prime}}} \nonumber \\
&&{\rm BR}(Z^\prime \to \Delta^{--} \Delta^{++}) ,
\label{Xsec1}
\eea
where $f_{q(\bar{q})}$ being the parton distribution function (PDF) from CTEQ6L \cite{Pumplin:2002vw} for (anti)quark. 
We show the variation of the cross section as a function of $M_{Z^\prime}$ at $\sqrt{s}=14$ TeV 
for $m_{\Delta^{\pm \pm}}=1.03$ TeV in Fig.~\ref{fig:Xsec-1}.
We find that the total BSM cross section within the invariant mass windows $M_{Z^\prime}\pm200$ GeV and $M_{Z^\prime}\pm250$ GeV for $M_{Z^\prime}=4$(5) TeV are 0.36(0.07) fb, 0.69(0.14) fb and 2.35 (0.5) fb for $x_H=-1$, 0, 1, respectively. 
In addition, we calculate the $\Delta^{\pm \pm}$ pair production cross section at the LHC 
from the purely SM process mediated by $Z$ and photon to be 
\bea
\sigma_{\rm SM}^{\rm LHC}&=& \int_{2 m_{\Delta}}^{E_{\rm LHC}} dM_{\rm inv}  \int_{\frac{M_{\rm inv}^2}{ E_{\rm LHC}^2}}^{1} dx \frac{4 M_{\rm inv}}{x E_{\rm LHC}^2}
\nonumber \\
&& \sum_{q, \overline{q}}\Big[ f_q (x, M_{\rm inv}) f_{\bar{q}} (\frac{M_{\rm inv}^2}{x E_{\rm LHC}^2}, M_{\rm inv})\Big] \hat{\sigma} , 
\label{SM-X}
\eea
where 
\bea
\hat{\sigma}&=& \frac{3.89\times 10^{8}}{144 \pi s} \Big(1-4 \frac{m_{\Delta^{\pm \pm}}^2}{s}\Big)^{\frac{3}{2}} s^2 \Big(A^2 Q_{u(d)}^2 e^2 
+ B^2(C_{V_{u(d)}}^2 \nonumber \\
&&+ C_{A_{u(d)}}^2)+ 2 A B e Q_{u(d)} C_{V_{u(d)}}\Big)
\eea
with $A= \frac{2 e}{s}$, $B= \frac{g_{Z} (1 - 2 \sin^{2}\theta_{W})}{s - m_{Z}^2}$, $e=\sqrt{\frac{4\pi}{128}}$, $Q_{u(d)}=\frac{2}{3}(-\frac{1}{3})$, $C_{V_{u(d)}}= g_Z ((-)\frac{1}{4}-Q_{u(d)}\sin^2\theta_W)$, $C_{A_u(d)}=g_Z(-\frac{1}{4})$, and $g_Z=2\frac{m_Z}{v}$, respectively. 
Here, $m_Z=91.2$ GeV denotes the $Z$ boson mass, and $v=246$ GeV represents the electroweak VEV. 
Integrating Eq.~(\ref{SM-X}) over the ranges 3800 GeV $\leq M_{\rm inv} \leq 4200$ GeV and 4750 GeV $\leq M_{\rm inv} \leq 5250$ GeV,  
we find the cross sections as $1.5\times 10^{-3}$ fb and $2.6 \times 10^{-4}$ fb, respectively, which are negligibly small compared with that
from $Z^\prime$ mediation.

Using Eq.~(\ref{Xsec1}), we also calculate the production cross sections of $\Delta^{\pm \pm}$ from $Z^\prime$ 
for $M_{Z^\prime}=4$ TeV, 5 TeV and 10 TeV, which are normalized by $g_X^2$, at 100 TeV proton proton future circular collider (FCC). 
Applying the invariant mass window cut $M_{Z^\prime}\pm200$ GeV (for $M_{Z^\prime}=4$ TeV) and $M_{Z^\prime}\pm250$ GeV (for $M_{Z^\prime}=5$ TeV and 10 TeV)
and $x_H=-1$ we find the cross sections as 116.4 fb, 73.25 fb and 5 fb respectively. We also find that the cross sections for $x_H=0$ the respective $Z^\prime$ masses are 190 fb, 118 fb and 30 fb
whereas that for $x_H=1$, the respective cross sections are 567 fb, 358 fb and 94 fb.
The corresponding cross sections are shown in the right panel of Fig.~\ref{fig:Xsec-1}. 
Using Eq.~(\ref{SM-X}), corresponding $\Delta^{\pm \pm}$ production cross sections from the SM gauge boson mediated processes 
are found to be 0.4 (0.204) fb, which are again negligibly small.

\begin{figure*}
\centering
\includegraphics[scale=0.28]{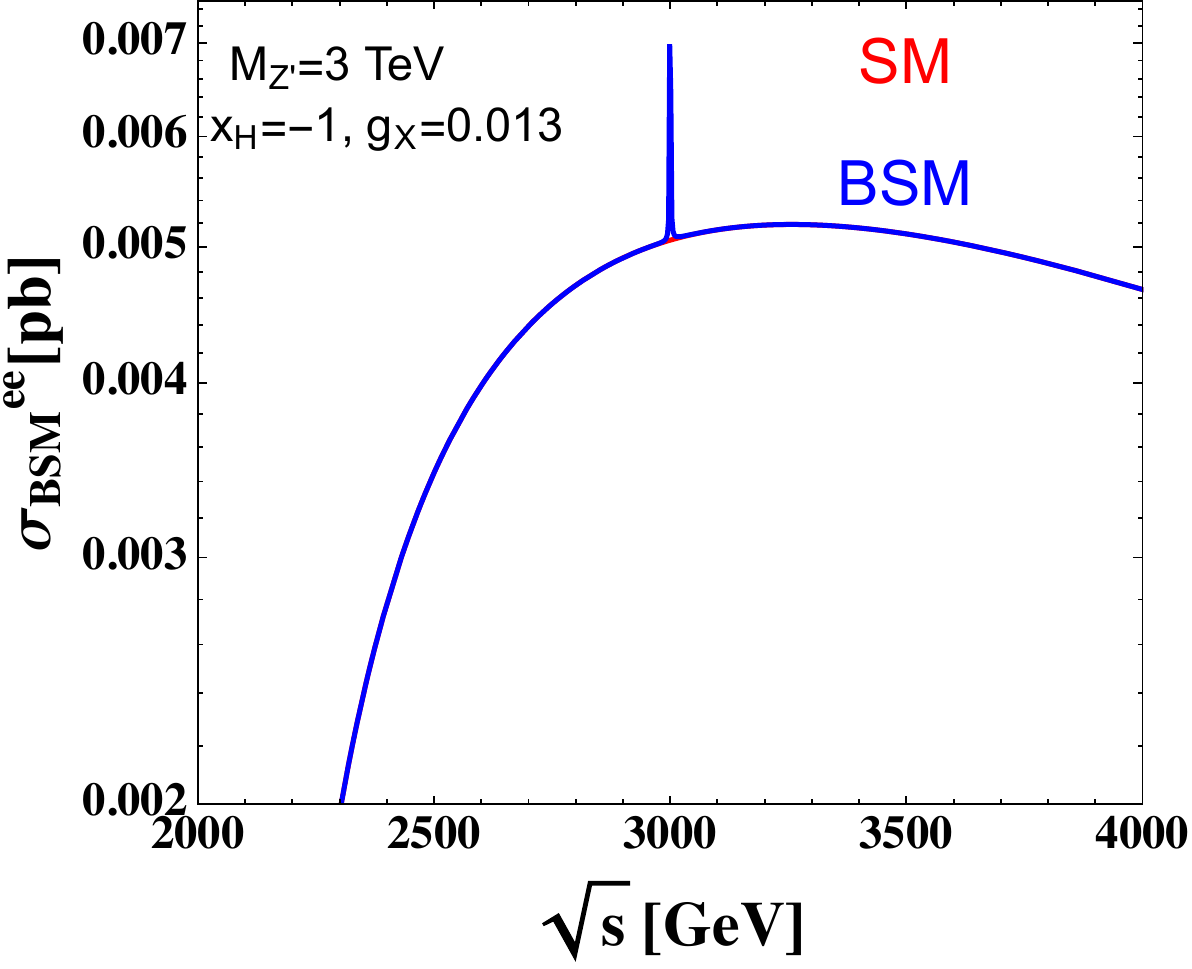}
\includegraphics[scale=0.28]{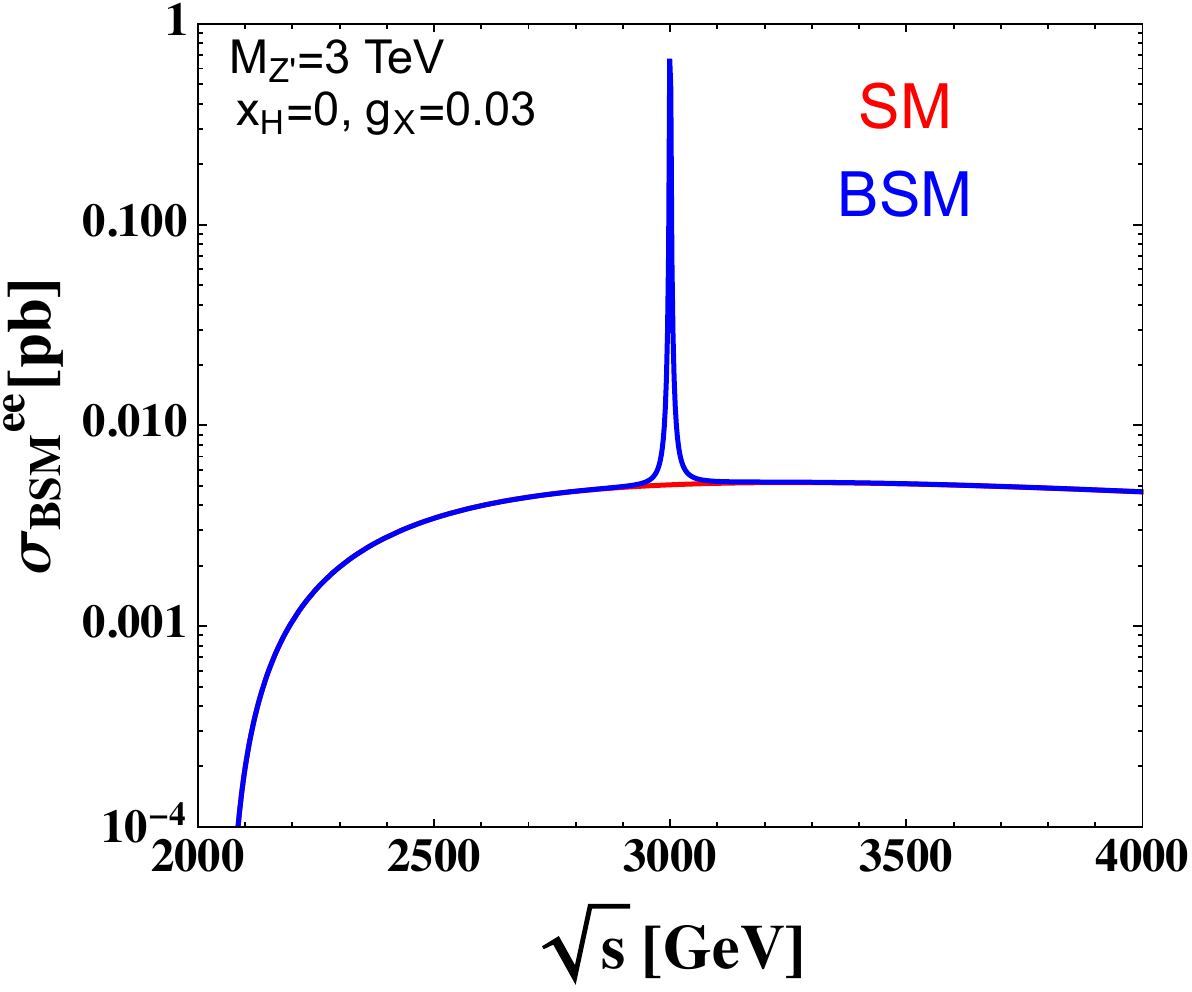}
\includegraphics[scale=0.28]{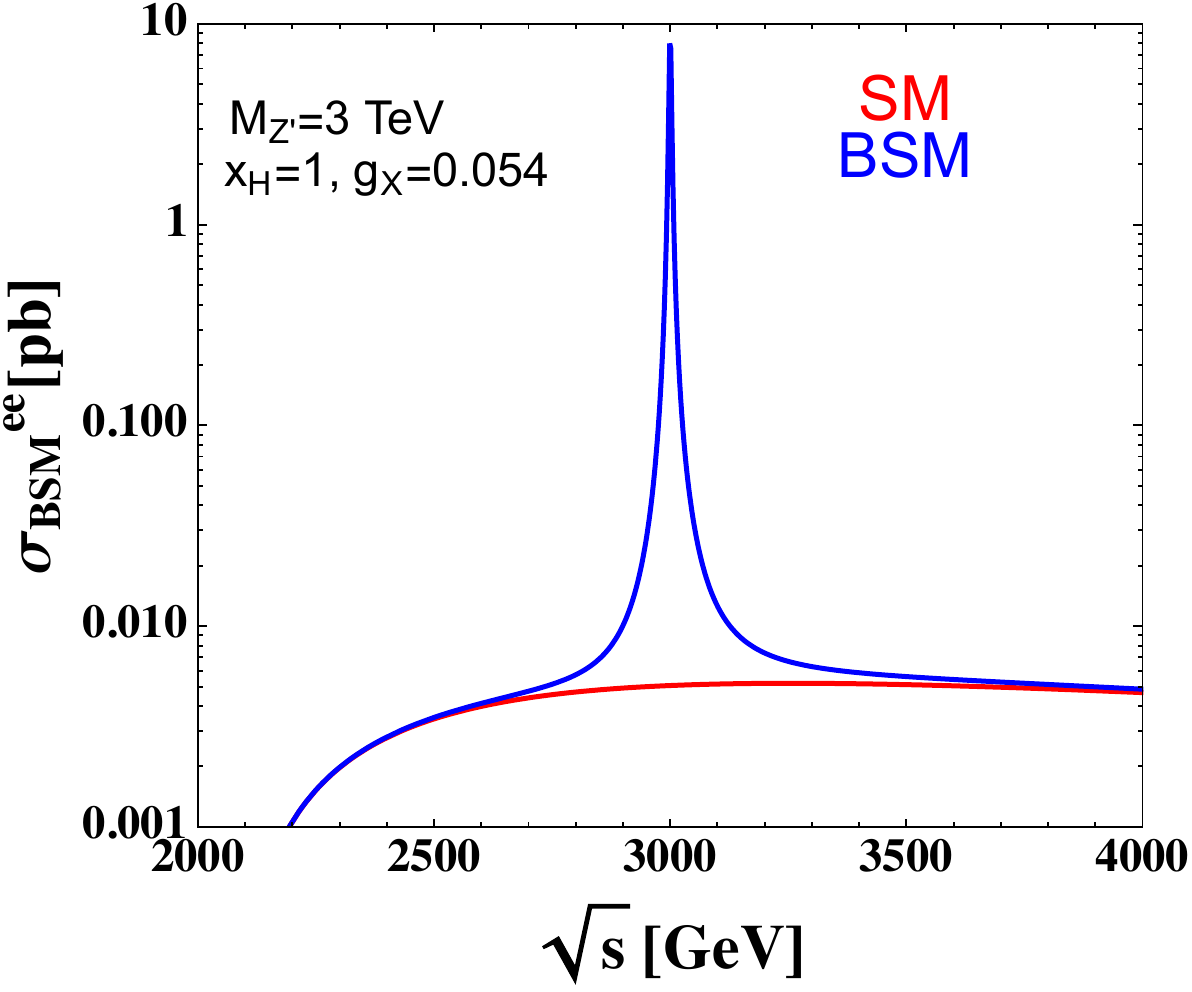}
\includegraphics[scale=0.28]{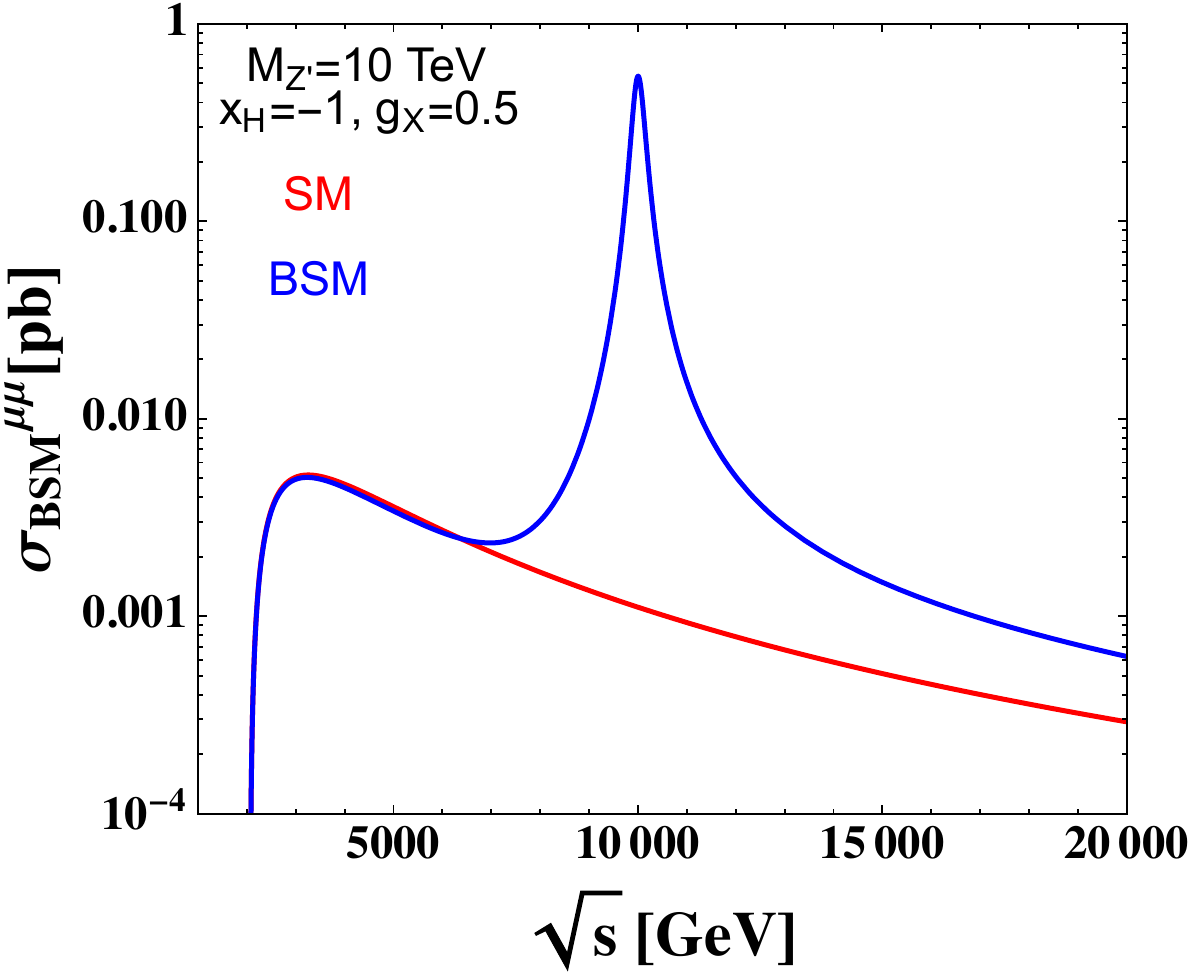}
\includegraphics[scale=0.28]{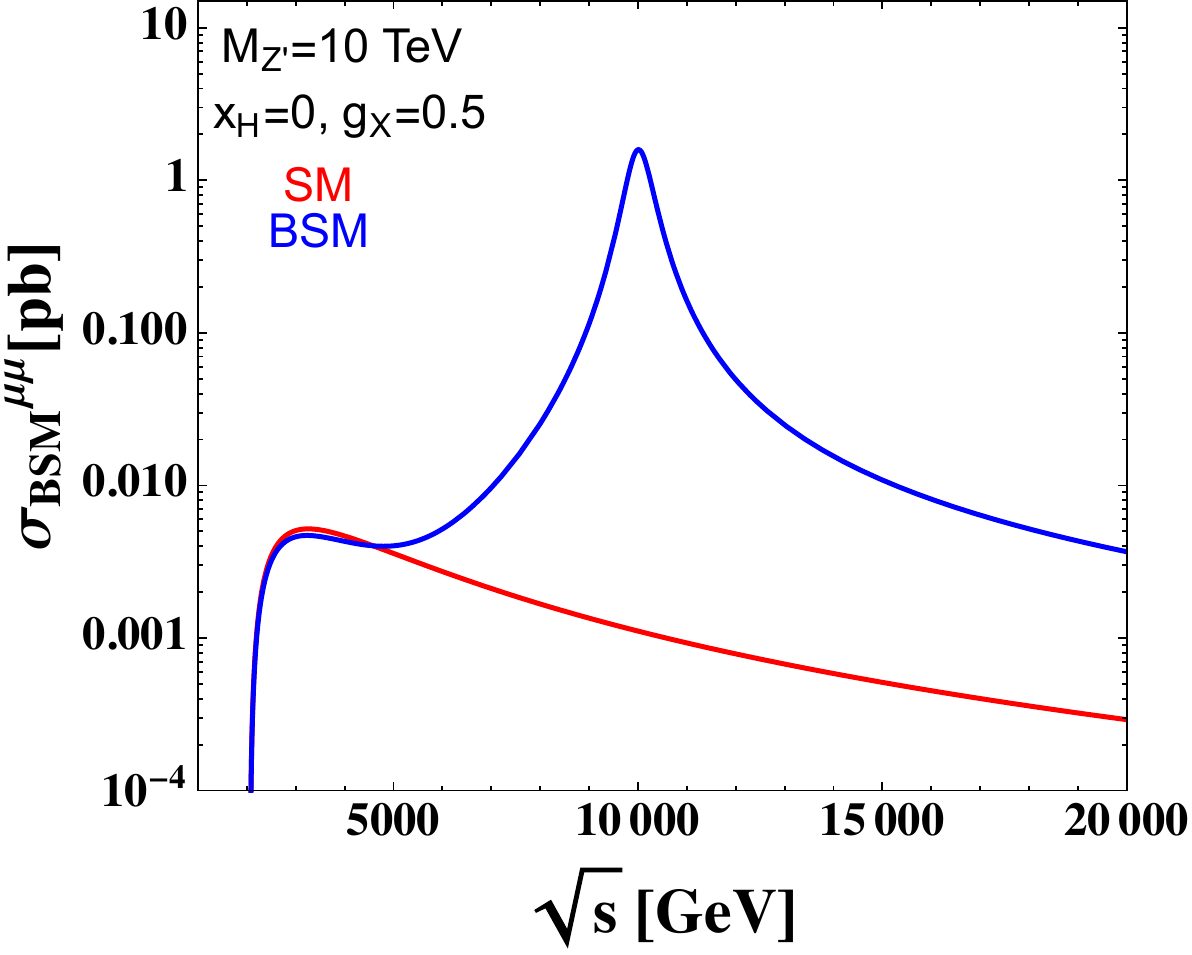}
\includegraphics[scale=0.28]{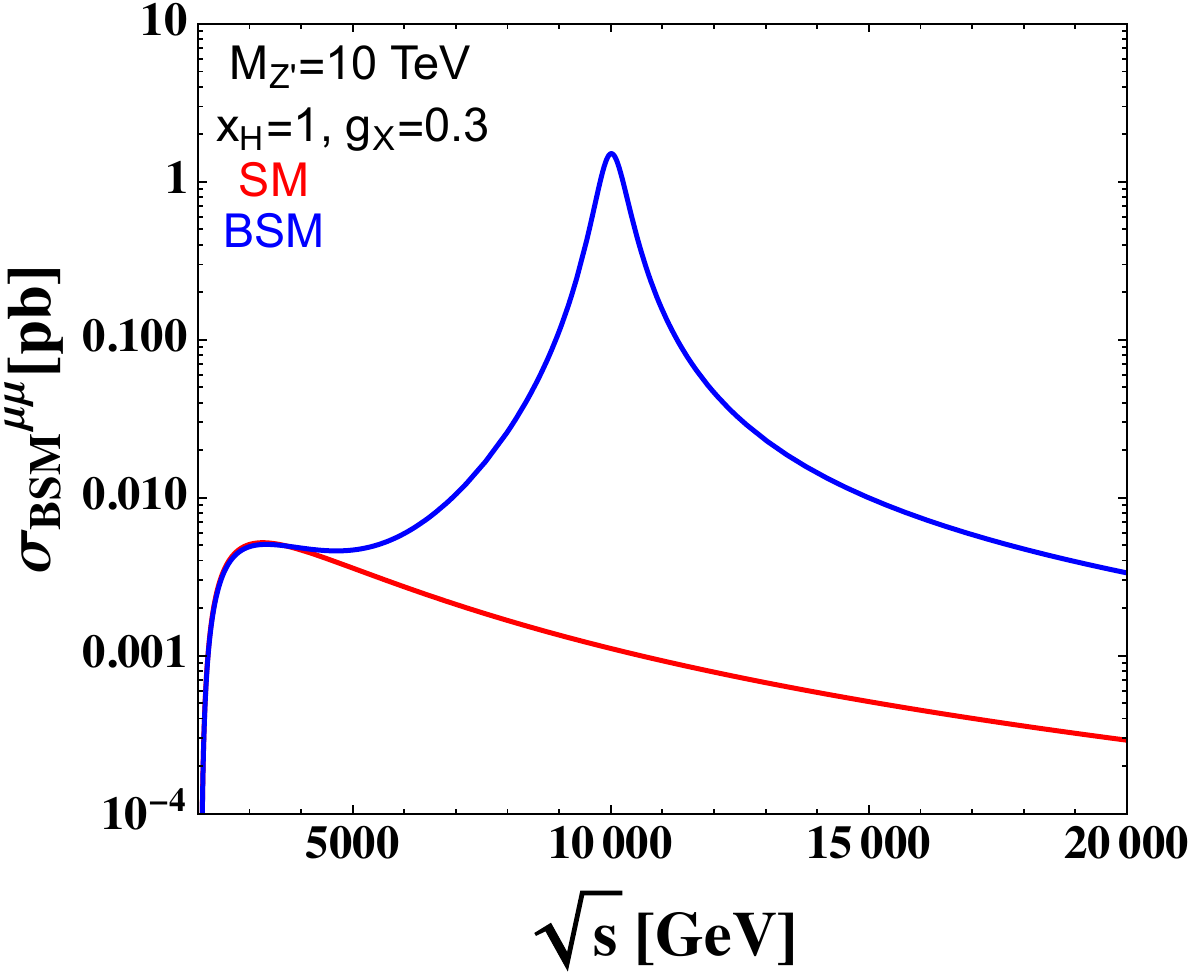}
\caption{$\Delta^{\pm\pm}$ production in $e^-e^+$ $(\mu^- \mu^+)$ colliders for different $x_H$ from SM $(Z, \rm{photon})$ and BSM $(Z^\prime)$ induced scenarios in the upper (lower) panel under $U(1)_X$ framework.}
\label{fig:Xsec-2-1}
\end{figure*}

We calculate the production cross section for $e^- e^+ \to \Delta^{++}\Delta^{--}$ process as
\bea
\sigma_{\rm BSM}^{ee/\mu \mu}&=& \frac{3.89\times 10^{8}}{48\pi s} (1 - 4 \frac{m_{\Delta}^2}{s})^{\frac{3}{2}} s^2 [\{A^2 {Q_{e}}^2{e}^2 +
 B^2 ({C_{V_{e}}}^2 + \nonumber \\
 && {C_{A_{e}}}^2)+ 2 A B e Q_{e} C_{V_{e}}\}+ 2\Re(B_1)\times \{A Q_{e}{e}^{2} Q_{V_{e}} + \nonumber \\
 &&B (C_{V_{e}}Q_{V_{e}}+ C_{A_{e}}Q_{A_{e}})\} +\nonumber \\
 &&\{\frac{g_{X}^2 (x_{H} + 2)^2 ({Q_{V_{e}}}^2 + {Q_{A_{e}}}^2)}{(s - M_{Z'}^2)^2 + M_{Z^\prime}^2 \Gamma_{Z^\prime}^2}\}].  
\label{Xsec2}
\eea
Our results are shown in the upper panel of Fig.~\ref{fig:Xsec-2-1} as a function of $\sqrt{s}$ for $x_H=-1, 0, +1$, respectively, from left to right.  
In the panels, purely the SM contributions mediated by $Z$ and photon are shown in red, 
while the cross sections including $Z^\prime$ mediated process are shown in blue, 
which exhibit sharp resonance peaks at $M_{Z^\prime}=3$ TeV as expected. 
At the $Z^\prime$ boson resonance point, we find that $Z^\prime$ induced $\Delta^{\pm \pm}$ production cross sections 
are 7 fb, 0.69 pb and 7.5 pb for $x_H=-1$, 0 and 1 for $g_X=0.013$, 0.03 and 0.054, respectively. 
The production cross section mediated by the SM gauge interactions is found to be 5 fb for $x_H=-1$, 
where  we have used 
$B_1= \frac{g_{X}^2(x_{H} + 2)^2}{(s - M_{Z'}^2) + \sqrt{-1}m_{Z'}\Gamma_{Z'}}$, 
$Q_{e} = -1$, $C_{V_{e}} = g_{Z} (-\frac{1}{4} - Q_{e} \sin^{2}\theta_{W})$, $C_{A_{e}} = g_{Z} \frac{1}{4}$, $Q_{V_{e}}= (-\frac{3}{4} x_{H} - 1) g_{X}$, and $Q_{A_{e}}= -\frac{1}{2} x_{H}  g_{X}$.

Using the Eq.~(\ref{Xsec2}) we calculate $\mu^+ \mu^- \to \Delta^{++} \Delta^{--}$ process and our results are shown in the lower panel 
of Fig.~\ref{fig:Xsec-2-1} as a function of $\sqrt{s}$ for $x_H=-1$ and 0 taking $g_X=0.5$ and for $x_H=1$ taking $g_X=0.3$, respectively, from left to right. 
SM contributions mediated by $Z$ and photon are shown in red, 
while the cross sections including $Z^\prime$ mediated process are shown in blue, 
which exhibit sharp resonance peaks at $M_{Z^\prime}=10$ TeV as expected. 
The width of the resonance depends on the size of the $U(1)_X$ coupling.
We consider the values of $g_X$ from Fig.~\ref{fig:limZp}.
The corresponding cross sections are approximately 600 fb, 1.76 pb and 1.6 pb respectively, for $x_H=-1$, 0 and 1 respectively.


We study decay of the doubly charged scalars into same sign dilepton $(\ell^{\pm} \ell^{\pm})$ 
and same sign gauge boson $(W^\pm W^\pm)$ from Eq.~(\ref{decayXX}). 
We focus on the $\ell^{\pm}\ell^{\pm}$ mode due to its dominance over $W^{\pm}W^{\pm}$ mode for $m_{\Delta^{\pm\pm}}=1.03$ TeV and $v_\Delta < 10^{-4}$ GeV. 
From Eq.~(\ref{Yuk1}), the light neutrino Majorana mass can be generated through the Yukawa coupling $Y^{ij}$ 
once the $SU(2)_L$ triplet scalar develops its VEV, 
$\mathcal{L}^{Y} \supset m_\nu \bar{\nu_L^c} \nu_L$, 
where $m_\nu=\frac{Y^{ij}}{\sqrt{2}} v_{\Delta}$ is  a $3\times 3$ neutrino mass matrix. 
This mass matrix is diagonalized as 
$U_{\rm MNS}^T m_\nu U_{\rm MNS}=diag(m_1, m_2, m_3)$ 
by the $3\times3$ unitary matrix, $U_{\rm MNS}$, which is a function of neutrino mixing angles $\theta_{12}= \frac{\sin^{-1}{\sqrt{0.87}}}{2}$, $\theta_{23} = \frac{\sin^{-1}{\sqrt{1}}}{2}$, $\theta_{13}= \frac{\sin^{-1}{\sqrt{0.092}}}{2}$ and Dirac CP phase $\delta_{CP}=\frac{3\pi}{2}$ \cite{ParticleDataGroup:2020ssz}.
\begin{figure*}
\centering
\includegraphics[scale=0.4]{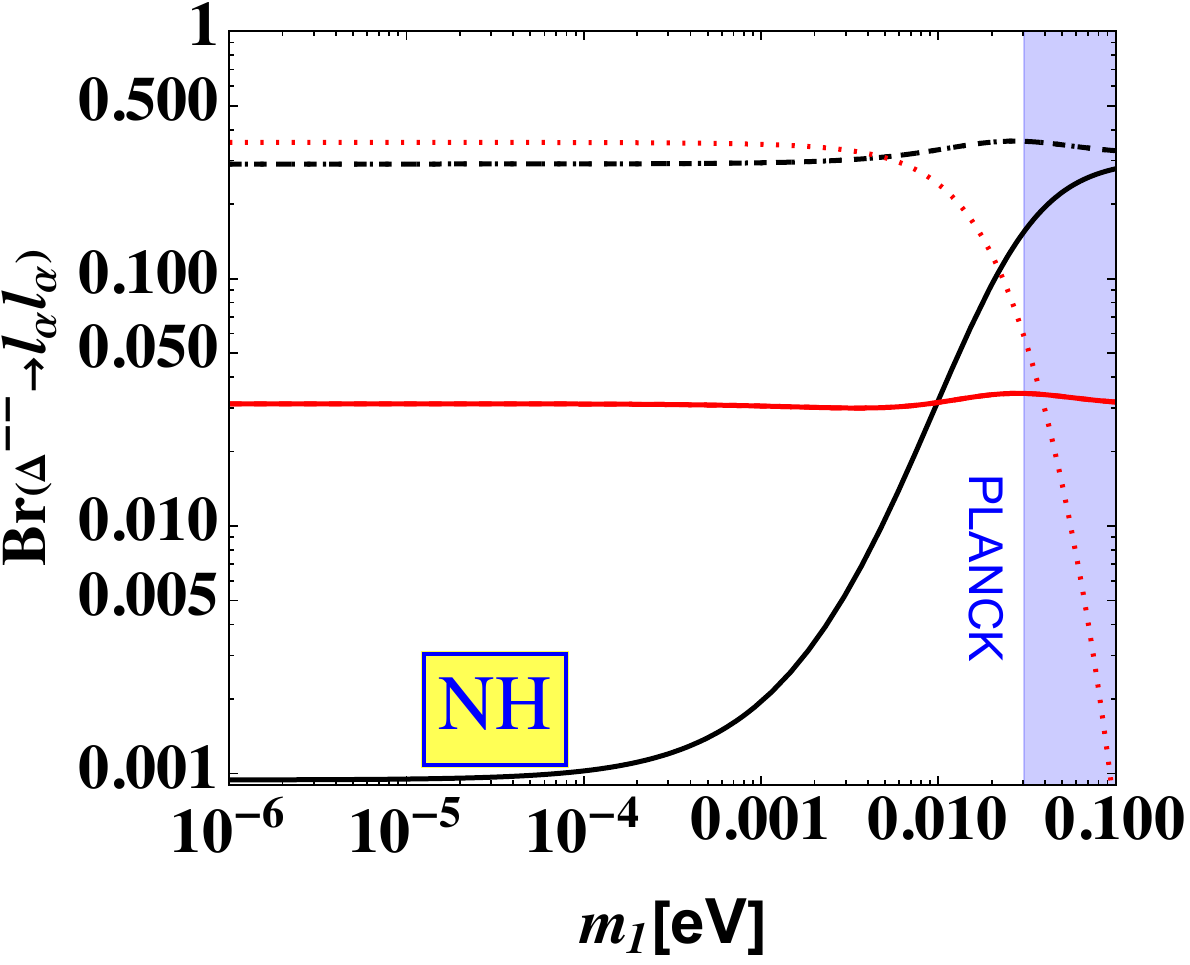}
\includegraphics[scale=0.4]{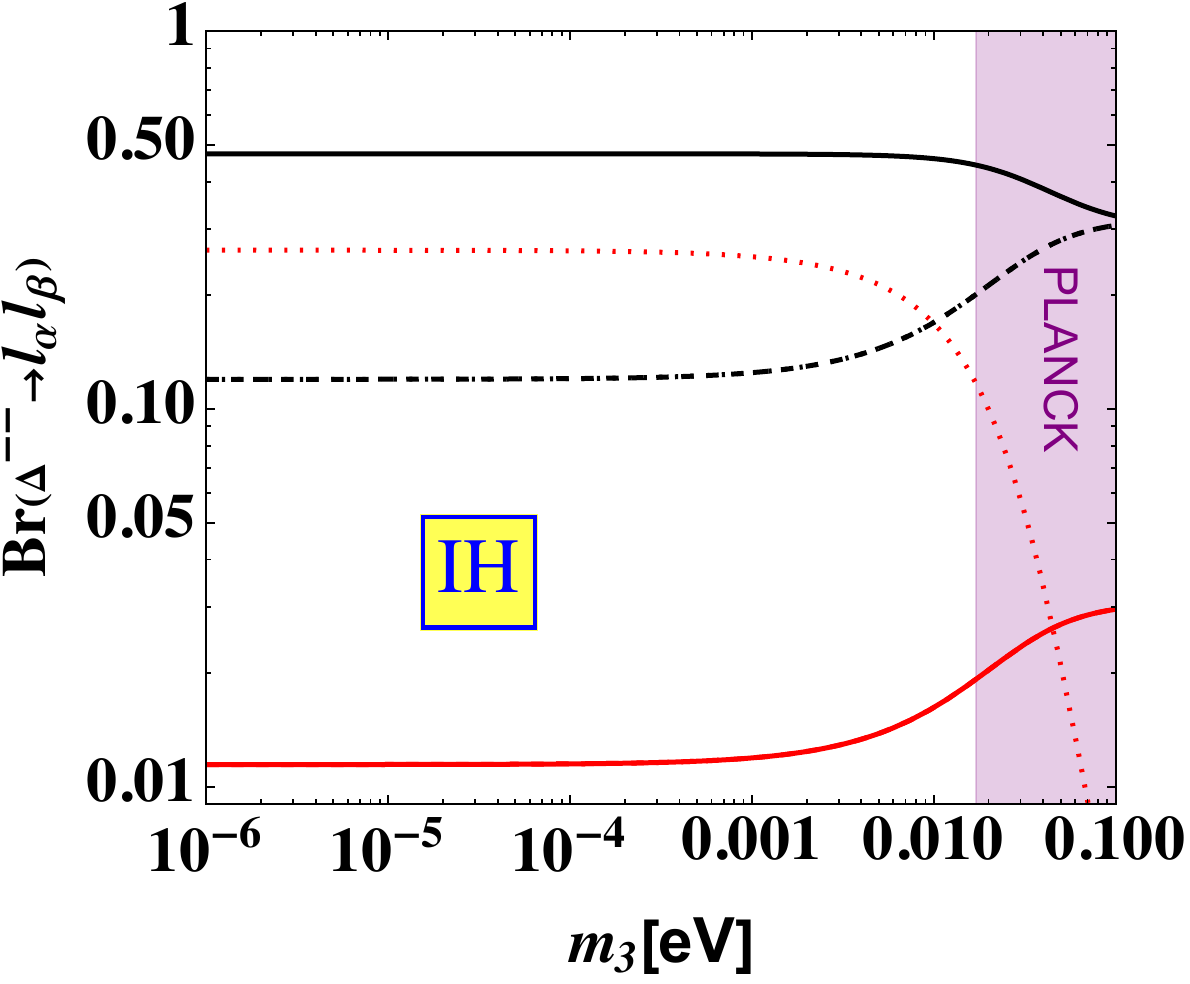}
\caption{Branching ratio of $\Delta^{--}$ into a pair of same sign, same (black) and different (red) flavor dileptons for the NH (IH) case in upper(lower) panel as a function of lightest light neutrino mass $m_{1(3)}$ for $m_{\Delta^{\pm \pm}}= 1.03$ TeV. 
In the NH and IH cases of same sign same flavor modes $(\Delta^{--}\to \ell_i^- \ell_i^-)$ branching ratios for $\Delta^{--} \to \mu^-\mu^-$ (dashed) and $\Delta^{--} \to \tau\tau$ (dotted) coincide with each other whereas $\Delta^{--}\to e^- e^-$ (solid) does not. 
In case of same sign different flavor modes $(\Delta^{--}\to \ell_i^- \ell_j^-)$ branching ratios for $\Delta^{--} \to e \mu$ (solid) 
and $\Delta^{--} \to e \tau$ (dashed) coincide with each other whereas $\Delta^{--} \to \mu^- \tau^-$ (dotted) does not. 
Shaded regions are excluded by PLANCK data for the NH (blue) and IH (purple) cases.}
\label{fig:BRtrip1}
\end{figure*}
The neutrino mass eigenvalues are determined by the neutrino oscillation data (mass squared differences), 
$\Delta m_{12}^2=m_2^2 -m_1^2 = 7.6\times10^{-5}$ eV$^{-2}$ and 
$|\Delta m_{23}^2|=|m_2^2 -m_1^2| = 2.4\times10^{-3}$ eV$^{-2}$, 
for two possible orderings for the mass eigenvalues: 
either $m_1 < m_2 < m_3$ the normal hierarchy (NH) or $m_3 < m_1 < m_2$ the inverted hierarchy (IH). 
For each case, the lightest mass eigenvalue, $m_1$ in NH while $m_3$ in IH, is left as a free parameter.  
For the NH case, the mass eigenvalue matrix is written as $diag(m_1, \sqrt{m_{12}^2+{m_{1}}^2_{\rm NH}}, \sqrt{m_{23}^2+{m_{2}}^2_{\rm NH}})$ whereas the one for the IH case as $diag(\sqrt{{m_{2}}^2_{\rm NH}-m_{12}^2}),\sqrt{m_{23}^2+{m_{3}}^2_{\rm NH}}, m_2)$.
Since the neutrino mass matrix is proportional to $Y^{ij}$, 
the total decay width of the doubly charged scalar is expressed as 
\bea
&\Gamma_{\rm Total}^{\rm{NH(IH)}} & = \nonumber\\
&C& \left(
 \sum_{i=1}^{3}{|[m_{\rm{NH(IH)}_\nu}]_{ii}|^{2} }+
  2\sum_{i<j=1}^{3}{|[m_{\rm{NH(IH)}_\nu}]_{ij}|^{2}} \right),~~~~~~
\eea 
where 
$C$ is a constant with mass dimension $-1$, 
$m_{\rm{NH(IH)}_\nu}$ is the neutrino mass matix for the NH (IH) case, and 
the index $i=1, 2, 3$ corresponds to the lepton flavor, $e, \mu, \tau$. 
Hence we can express the branching ratio of the doubly charged scalar into a pair of same sign different flavor leptons as
\bea
\rm{BR}_{\rm{NH(IH)}}^{\rm{off-diag}}&=&\frac{ 2 \sum_{i<j=1}^{3}{|[m_{\rm{NH(IH)}_\nu}]_{ij}|^{2}}}{\Gamma_{\rm{Total}}^{\rm{NH(IH)}}/C} \nonumber \\
&=&\rm BR(\Delta^{--} \to \ell_\alpha \ell_\beta)~
\eea
and the branching ratio of the doubly charged scalar into a pair of same sign same flavor lepton as 
\bea
\rm{BR}_{\rm NH(IH)}^{\rm diag}&=& \frac{\sum_{i=1}^{3}{|[m_{\rm{NH(IH)}_\nu}]_{ii}|^{2} }}{\Gamma_{\rm Total}^{\rm NH(IH)}/C} \nonumber \\
&=&\rm BR(\Delta^{--} \to \ell_\alpha \ell_\alpha),~~~~~~~
\eea
respectively.

We show the branching ratios of $\Delta^{--}$ into a pair of same sign lepton in the upper (lower) panel of Fig.~\ref{fig:BRtrip1} 
for the NH (IH) case for the same and different flavors as a function of lightest neutrino mass eigenvalue $m_{1(3)}$. 
\begin{figure*}
\centering
\includegraphics[scale=0.4]{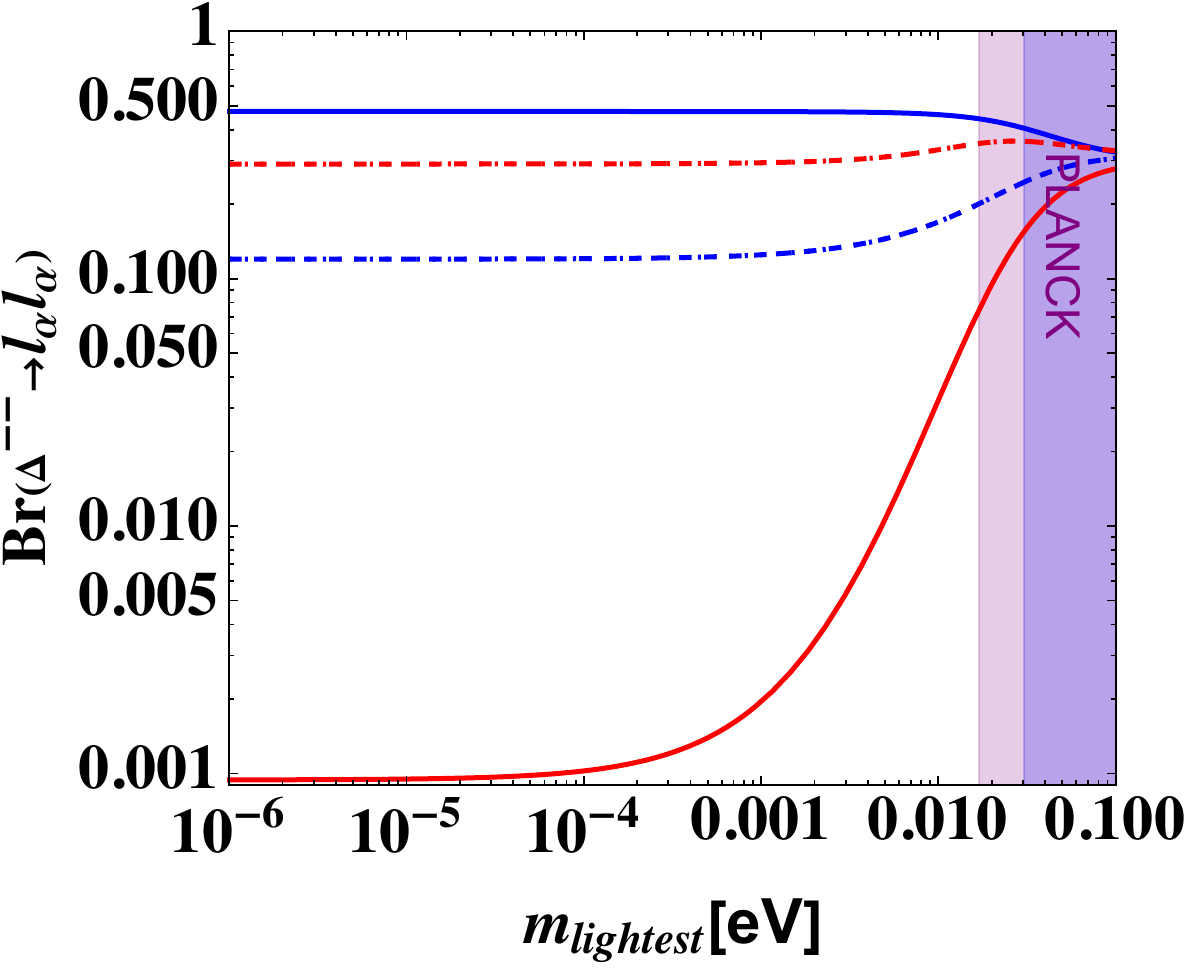}
\includegraphics[scale=0.4]{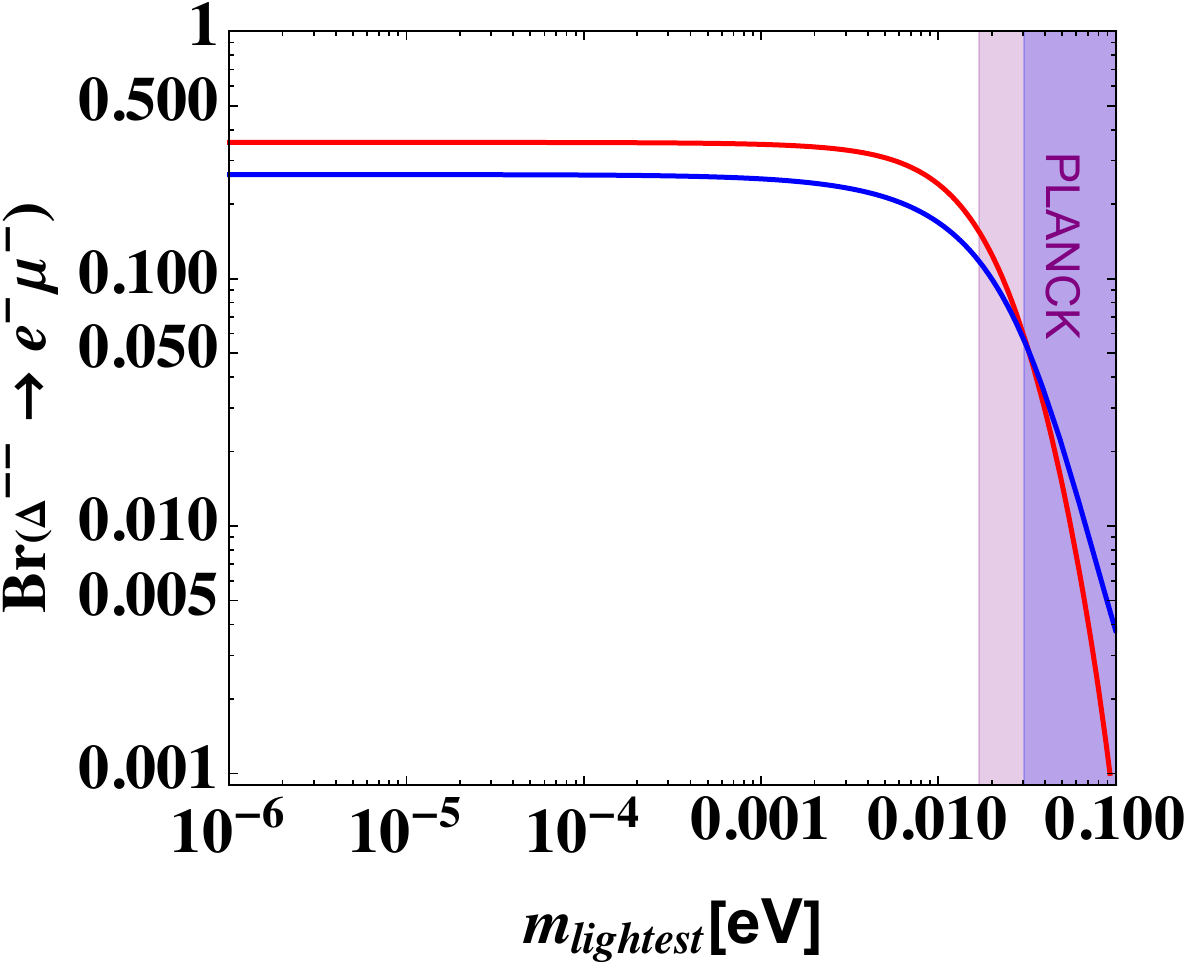}
\caption{Branching ratio of $\Delta^{--}$ into a pair of same sign same (different) flavor dileptons for the NH (red) and IH (blue) case in upper(lower) panel as a function of lightest light neutrino mass for $m_{\Delta^{\pm \pm}}=1.03$ TeV. 
In case of same sign same flavor modes $(\Delta^{--}\to \ell_i^- \ell_i^-)$ branching ratios for $\Delta^{--} \to \mu^-\mu^-$ (dashed) and $\Delta^{--} \to \tau^-\tau^-$ (dotted) coincide with each other whereas $\Delta^{--} \to e^- e^-$ (solid) does not. 
Decay of $\Delta^{--}$ into same sign different flavor modes are represented by the branching ratio of $\Delta^{--} \to e^-\mu^-$ in the lower panel for NH (red) and IH (blue) cases. Shaded regions are excluded by PLANCK data for the NH (blue) and IH (purple) cases.}
\label{fig:BRtrip2}
\end{figure*}
In NH case branching ratio of $\Delta^{--}$ into same sign same flavor dilepton are shown by black solid $(ee)$, dashed ($\mu \mu$) and dotted $(\tau \tau)$ lines where $\mu \mu$ and $\tau \tau$ cases coincide with each other. We find that the branching ratio into $ee$ mode increases with $m_1$ whereas those of $\mu\mu$ and $\tau\tau$ modes remain almost same over the range of $m_1$ under consideration. In this case we consider the PLANCK limit on the lightest neutrino mass eigenvalue as $m_1 > 0.03026$ eV is ruled out in NH case \cite{Planck:2018vyg} and it is shown by blue shaded region satisfying the upper limit on the sum of the neutrino mass eigenvalues $\sum_i m_{i} < 0.12$ eV. On the other hand we find branching ratios of $\Delta^{--}$ in different flavors involving $e\mu$ and $e \tau$ modes being denoted by red solid and dashed lines coincide with each other and remain almost same through out the range of $m_1$ under consideration except near the vicinity of the PLANCK exclusion limit $m_{1}> 0.003$ eV throughout the mass range of $m_1$, however, $\mu \tau$ mode remains almost constant which is denoted by red dotted line. 

In the lower panel of Fig.~\ref{fig:BRtrip1} we show the IH case where same sign same flavor cases by solid black $(ee)$, dashed $(\mu\mu)$ and dotted $(\tau \tau)$ lines, respectively where $\mu \mu$ and $\tau\tau$ modes coincide with each other. Same sign different flavor modes are shown by the red solid $(e\mu)$, dashed $(e\tau)$ and dotted $(\mu \tau)$ lines, respectively where $e\mu$ and $e\tau$ modes coincide. These different flavor scenarios evolve from the off-diagonal entries of $Y^{ij}$. In this case branching ratios of respective modes remain almost constant at their respective values up to the vicinity of the PLANCK exclusion limit $m_3 > 0.01701$ eV which is represented by the shaded region in purple.  

In the upper panel of Fig.~\ref{fig:BRtrip2} we compare branching ratio of same sign same flavor dilepton modes $ee$, $\mu \mu$ and $\tau \tau$ from NH (IH) represented by red (blue) solid, dashed and dotted lines respectively with the PLANCK exclusion limits for the NH(IH) case shown by blue(purple) shaded regions. We find that for NH and IH cases $\mu\mu$ and $\tau\tau$ modes coincides. It is found that  branching ratios of $ee$, $\mu\mu$ and $\tau\tau$ modes vary between $12\%-48.3\%$ when lightest neutrino mass eigenvalue is lower than the PLANCK exclusion limit for NH and IH case. Branching ratio of $ee$ mode from NH case is below $1\%$ for $m_{\rm lightest} < 0.0047$ eV, however, it reaches up to $16\%$ for $m_{\rm lightest}=0.03026$ eV, that is, at the boundary of the PLANCK exclusion limit. 
The branching ratios of $\mu\mu$ and $\tau\tau$ remain almost fixed at $30\%$ for for $m_1$ below PLANCK exclusion limit and $34.4\%$ for $m_1=0.01$ eV. 
We show the leading $\Delta^{--}\to e^- \mu^-$ mode for the NH (IH) case in the lower panel by red (blue), 
where the corresponding branching ratio decreases at the vicinity of the PLANCK exclusion limit, however, before that it remains constant at $36(27)\%$ for $m_{\rm lightest}< \mathcal{O}(0.001)$ eV. 
Considering three generations of neutrinos with non-zero light neutrino mass eigenvalues satisfying PLANCK exclusion limits, 
we can calculate the number of events of multilepton signals from $\Delta^{\pm \pm}$ pair produced at pp, $e^- e^+$ and $\mu^-\mu^+$ colliders, respectively. 
Further we can compare these events with those obtained from the SM backgrounds to estimate prospective signal significance 
for the NH and IH cases. 

From Fig.~\ref{fig:BRtrip1}, considering the NH case and $m_{1}=0.01 (0.001)$ eV, we find  branching ratios of $\Delta^{--}$ into same sign dilepton modes as: (i) BR$(ee)$: $3.0(0.21)\%$, (ii) BR$(\mu\mu/ \tau \tau)$ $34.0(30.0\%)$, (iii) BR$(e\mu/e\tau)$ $3.0(3.0)\%$ 
and (iv) BR$(\mu\tau)$: $23.0(33.79)\%$, respectively. 
For the IH case with $m_{3}=0.01 (0.001)$ eV, we find dilepton branching ratios of $\Delta^{--}$ into same sign dilepton modes as: (i) BR$(ee)$: $46.0(48.0)\%$, (ii) BR$(\mu\mu/ \tau \tau)$: $17.0(12.3\%)$, (iii) BR$(e\mu/e\tau)$ $1.5(1.2)\%$ and (iv) BR$(\mu\tau)$: $17.0(25.5)\%$, respectively. 
We then use these branching ratios to estimate the number of signal events from $\Delta^{--}$ pair production 
at different colliders for both $Z^\prime$ and SM gauge boson mediated processes as
\bea
\rm{Events}_{\rm BSM/SM}^{NH/IH}&=&\sigma_{\rm BSM/ SM}^{\rm LHC/e^-e^+/\mu^-\mu^-}\times  \nonumber \\
&&\rm(BR(\Delta^{--}\to \ell \ell)^{\rm NH/IH})^2 \times \mathcal{L}, 
\label{evt}
\eea
where $\mathcal{L}$ is the luminosity of colliders such as 3 ab$^{-1}$ (LHC) and 1 ab$^{-1}$ $(e^-e^+/ \mu^- \mu^+)$ and $\sigma_{\rm BSM/ SM}^{\rm LHC/e^-e^+/\mu^-\mu^-}$ will be obtained from Eqs.~(\ref{Xsec1}) and (\ref{Xsec2}), respectively. 
Noting the branching ratio of $\Delta^{\pm \pm}$ into $2\mu$ $(2e)$ mode is large compared to the other modes in the NH (IH) case,
we focus on this mode to obtain as many events as possible. 
We do not consider the final states involving $\tau$ leptons due to its further decay into different visible modes 
which further reduces the number of events. 

The production cross section of $\Delta^{\pm \pm}$ in the LHC are shown in the left panel of Fig.~\ref{fig:Xsec-1}.  After $\Delta^{\pm\pm}$ production at the LHC we estimate the signal events using Eq.~(\ref{Xsec1}) for $x_H=-1$, 
where the corresponding $U(1)_X$ couplings are obtained from Fig.~\ref{fig:limZp} as $g_X=0.18$ and $0.38$ 
for $M_{Z^\prime}=4$ TeV and 5 TeV for $M_{Z^\prime}\pm200$ GeV and $M_{Z^\prime}\pm250$ GeV for $M_{Z^\prime}=4$(5) TeV. Considering the leading decay modes of $\Delta^{--}$ and using Eq.~(\ref{evt}), we estimate the NH case with $m_{1}=0.01 (0.001)$ eV
and obtain the $4\mu$ final state with 5 (4) events for $M_{Z^\prime}=4$ TeV and 3.5(2.8) events for $M_{Z^\prime}=5$ TeV, respectively. 
In the NH case, $4e$ events from $Z^\prime$ mediated process will be about two orders of magnitude smaller than the $4\mu$ process for $m_1=0.01$ eV, and that for $m_1=0.001$ will be negligibly small following the smallness of BR$(\Delta^{--}\to e^- e^-)$. 
On the other hand for IH case with $m_{3}=0.01 (0.001)$ eV, we obtain the number of events for $4e$ final states 
as 9.1 (10) and 1.8 (2.0) for $M_{Z^\prime}=4$ TeV and 5 TeV, respectively. 
Simulating the generic SM backgrounds for four lepton final states from $pp\to \ell^+ \ell^- \ell^+ \ell^-$ using  MadGraph \cite{Alwall:2011uj}, we find the cross sections for LHC at $\sqrt{s}=14$ TeV as $4.8 \times 10^{-3}$ pb for $\ell= e,~\mu$. 
Imposing invariant mass $(M_{\rm inv}^{\ell \ell})$ cuts for opposite sign dilepton of same flavor 980 GeV $\leq M_{\rm inv}^{\ell \ell}\leq$ 1.08 TeV, we find that the background cross section reduces to $\mathcal{O}(10^{-15})$ pb which is essentially zero.  
Thus, the significance of the $4\mu$ signals at LHC, defined as $\rm Significance= Signal/\sqrt{Signal+ SM~background}$, 
would be probed at 2.23 (2.0)$\sigma$ for $m_1=0.01$ (0.001) eV in the NH case for $M_{Z^\prime}=4$ TeV, 
whereas that for $M_{Z^\prime}=5$ TeV would be probed with less than 2$\sigma$ significance. 
In the IH case, we find that the significance of probing $4e$ final state can be around 3$\sigma$ for $M_{Z^\prime}=4$ TeV 
with $m_3=0.01$ eV and 0.001 eV whereas that for $M_{Z^\prime}=5$ TeV would be less than 2$\sigma$. 

We study $4\mu$ and $4e$ final states at the LHC for $x_H=0$ and $m_{1(3)}=0.01$ eV and 0.001 eV for the NH (IH) case.
Here we set  $g_X=0.1$ and $0.2$ and the invariant mass cut $M_{Z^\prime}\pm200$ GeV and $M_{Z^\prime}\pm250$ GeV for $M_{Z^\prime}=4$ and 5 TeV, respectively. 
We find that in the NH case for $m_{1}=0.01$ (0.001) eV there are 1.53(1.2) $4\mu$ events for $M_{Z^\prime}=4$ TeV whereas those for $M_{Z^\prime}=5$ TeV are also of the same values. 
For the IH case, we find that for $m_{1}=0.01$ (0.001) eV there are  4.4(4.8) $4e$ events for $M_{Z^\prime}=4$ TeV whereas those for $M_{Z^\prime}=5$ TeV are 3.6 (3.9). 
Estimating the SM backgrounds for four lepton final states from $pp\to \ell^+ \ell^- \ell^+ \ell^-$ by using MadGraph \cite{Alwall:2011uj}, 
we find the cross sections for LHC at $\sqrt{s}=14$ TeV as $4.8 \times 10^{-3}$ pb for $\ell= e,~\mu$. Imposing invariant mass $(M_{\rm inv}^{\ell \ell})$ cuts for opposite sign dilepton of same flavor 980 GeV $\leq M_{\rm inv}^{\ell \ell}\leq$ 1.08 TeV we find that the background cross section reduces to $\mathcal{O}(10^{-15})$ pb. 
Hence, we find that the significance to observe the $4\mu$ and $4e$ final states from $M_{Z^\prime}=4$ TeV and 5 TeV in the NH cases 
will be less than 2$\sigma$. 
However, in the IH case, $4e$ final state can be observed with a significance of 2$\sigma$ or slightly more than that for $m_3=0.01$ eV and $0.001$ eV from $M_{Z^\prime}=4$ TeV. 
Also,  $4e$ final state can be observed with a significance of 2$\sigma$ in the IH case for $m_3=0.001$ eV from  5 TeV at the LHC. 

We also study the significance of probing $4\mu$ ($4e$) final state at the LHC for $x_H=1$ with $m_{1(3)}=0.01$ eV and 0.001 eV 
for the NH (IH) case by using $g_X=0.043$ and $0.1$ and the invariant mass cut $M_{Z^\prime}\pm200$ GeV and $M_{Z^\prime}\pm250$ GeV for $M_{Z^\prime}=4$ and 5 TeV, respectively. 
We find that the number of events for $4\mu$ final states in the NH case will be 1.5 (1.2) for $m_1=0.01$ (0.001) eV and 1.73 (1.35) for $m_1=0.01$ (0.001) eV for $M_{Z^\prime}=4$ TeV and 5 TeV, respectively. 
We estimate number of events for $4e$ final states in the IH case as 2.76 (3.0) for $m_1=0.01$ (0.001) eV and 3.2 (3.5) for $m_1=0.01$ (0.001) eV for $M_{Z^\prime}=4$ TeV and 5 TeV, respectively.  
Hence, for $x_H=1$, the significance for both the NH and IH cases would be less than 2$\sigma$.

We study the production of $\Delta^{\pm \pm}$ at a 100 TeV proton-proton future circular collider (FCC) from $Z^\prime$ and the production cross sections are shown in the right panel of Fig.~\ref{fig:Xsec-1}. 
Using the same values of $g_X$ for $x_H=-1$, 0 and 1 and taking $M_{Z^\prime}=4$ TeV, we estimate the number of signal events for four leptons in the final state involving electrons and muons. 
We first calculate the SM backgrounds for $pp\to \ell^+ \ell^- \ell^+ \ell^-$ process ($\ell= e, \mu$) and find 0.04 pb for each flavor. 
Imposing invariant mass $(M_{\rm inv}^{\ell \ell})$ cuts for opposite sign dilepton of same flavor 980 GeV $\leq M_{\rm inv}^{\ell \ell}\leq$ 1.08 TeV, we find that the background cross section reduces to $0.072$ fb which provides 216 events for 3 ab$^{-1}$ luminosity. 
The production cross sections of $\Delta^{\pm \pm}$ are 116.4 fb, 190.0 fb and 567.3 fb for $x_H=-1,$ 0 and 1, respectively, 
within the invariant mass window $M_{Z^\prime}\pm200$ GeV. 
For 3 ab$^{-1}$ luminosity, the signal events for 4$e$ final state for $x_H=-1$ will be 314.28 for $m_1=0.01$ eV. 
Therefore, the significance of observing 4$e$ final state can be more than 5$\sigma$, however, that for  $m_1=0.001$ eV will be very small due to the tiny branching ratio of $\Delta^{\pm \pm}$ into $2e$. 
Similarly, we calculate the total number of signals for 4$\mu$ final state as $4.037 \times 10^{4}$ and $3.143\times 10^{4}$  for $m_1=0.01$ eV and $m_1=0.001$ eV, respectively, which can exceed $5\sigma$. 
Following the same line, we also estimate the number of signal events for $2e2\mu$ final state as 314.3 events 
which can also be probed with more than $5\sigma$ significance. 
In the IH case, $4e$ final state provides $7.39\times 10^{4}$ and $8.046\times 10^{4}$ events for $m_3=0.01$ eV and 0.001 eV, respectively, 
whose significance can exceed $5\sigma$ whereas $4\mu$ final state provides $1.01\times 10^{4}$ and $5.3\times 10^{3}$ events, 
which can also exceed $5\sigma$ significance. 
We estimate that $2e2\mu$ final state provides us with 79 and 50 events in the IH case for $m_3=0.01$ eV and 0.001 eV, respectively, 
which correspond to $4.6\sigma$ and $3\sigma$ significance. 
Note that $\Delta^{\pm \pm}$ production cross section and for $x_H=0$ and $1$ are more than that we obtained for $x_H=-1$. 
We now conclude that the significance of probing $4e$, $4\mu$ and $2e2\mu$ final states at the 100 TeV collider will be sizable if produced by the $Z^\prime$ mediated processes. 
In case of FCC at 100 TeV, if we consider $M_{Z^\prime}=10$ TeV, then $\Delta^{\pm \pm}$ production cross section
will be 10 fb, 16 fb and 48 fb for $x_H=-1$, $0$ and $1$, respectively. 
Within the invariant mass window $M_{Z^\prime}\pm200$ GeV, the cross sections become 4.89 fb, 30.06 fb and 94.4 fb for $x_H=-1,$0 and $1$, respectively. 
First we consider the case of $x_H=-1$ and find $1.7\times 10^{3}$ $(1.323 \times 10^{3})$ events for $4\mu$ final states considering $m_1=0.01$ $(0.001)$ eV in the NH case.  
This case can be observed with more than $5\sigma$ significance in 100 TeV collider. 
For the IH case,  we find  $\Delta^{\pm \pm} \to e^{\pm} e^{\pm}$ is $46.0\% $ $(48.0\%)$
and $\Delta^{\pm \pm} \to \mu^{\pm} \mu^{\pm}$ is $17.0\% $ $(12.3\%)$ for $m_3=0.01$ $(0.001)$ eV which will also produce more signal events 
than the NH case.
Therefore, in the same line, we predict that these signals can be observed in FCC at 100 TeV proton-proton collider with more than $5\sigma$ in future.
We find that $\Delta^{\pm \pm}$ production cross section for $x_H=0$ $(1)$ is 30.06 (94.4) fb which is roughly one (two) order of magnitude 
more than the case with $x_H=-1$.
Such an enhancement in signal events will be reflected in observing $4e$, $4\mu$ and $2e2\mu$ final states with more than $5\sigma$ significance, 
keeping the same numbers of SM background events. 
We thus conclude that  the 1.03 TeV doubly charged triplet production from 10 TeV $Z^\prime$ is interesting to study neutrino mass hierarchy 
at the future 100 TeV FCC.

From Fig.~\ref{fig:Xsec-2-1} and using Eq.~(\ref{evt}), we estimate signal events at $e^-e^+$ colliders for $x_H=1$ 
and $M_{Z^\prime}=3$ TeV. 
Setting 1 ab$^{-1}$ luminosity for $4\mu$ final state in the NH (IH) case as $8.67\times10^{5}$ $(2.1675\times 10^{5})$ for $m_1=0.01$ eV and it becomes $6.75\times 10^{5}$ $(1.135\times 10^{5})$ for $m_{1}=0.001$ eV. 
The corresponding number of events from the purely SM mediated processes is roughly $\mathcal{O}(3-4)$ of magnitude smaller 
than the $Z^\prime$ mediated processes. 
Due to the dependence of BR$(\Delta^{--}\to \ell^- \ell^-)$ on lightest light neutrino mass $m_{1(3)}=0.01$ eV we find number of events for $4e$ final state from NH (IH) cases is $6.75\times 10^{3} (1.6 \times 10^{6})$, however, that for $m_{1(3)}=0.001$ eV is 33$(1.73 \times 10^{6})$, respectively. Considering both the $\Delta^{--}$ decaying into $e^- \mu^-$ produce $6.75\times 10^{3}(1.7 \times 10^{3})$ events for NH(IH) case taking $m_{1(3)}=0.01$ eV whereas that for $m_{1(3)}=0.001$ eV is $6.75\times 10^{3}(1.08\times 10^{3})$, respectively. In case of $e^-e^+$ collider we generate the generic SM backgrounds from $e^- e^+ \to \ell^+ \ell^- \ell^+ \ell^-$ process with $\ell= e,~\mu$ at $\sqrt{s}=3$ TeV where the cross sections are $8\times 10^{-4}$ pb, $1.44\times 10^{-5}$ pb and $8.05\times 10^{-4}$ pb for $4e$, $4\mu$ and $2e 2\mu$ final states, respectively. For a luminosity of 1 ab$^{-1}$, total number of SM background events will be $\mathcal{O}(1-100)$. Due to a comparatively small number of SM background we find that $4e$, $4\mu$ and $2e2\mu$ final states could be observed with a significance of more than $5\sigma$ which is significantly dominant over the $4e$, $4\mu$ and $2e2\mu$ final states coming from purely SM processes. 
In this analysis we did not consider $4\tau$, $\mu\tau$ and $e\tau$ modes for simplicity as we are not analyzing $\tau$ decay which further reduces the significance, however, these modes could be studied in future using $\tau$ decay. In addition to that we find $\Delta^{\pm \pm}$ production cross section for $x_H=0$ is roughly one order of magnitude less than that we obtained in case of $x_H=1$ whereas that for $x_H=-1$ is three orders of magnitude less that the cross section obtained from $x_H=1$ which will be reflected in the corresponding signal events, however, that will be significant for $x_H=0$ case where 4$\mu$ final state could be observed with a significance of 5$\sigma$ whereas that for $x_H=-1$ could reduce to a significance of 2$\sigma$ depending on the neutrino mass hierarchy. Hence we predict that depending on the neutrino mass hierarchy $\Delta^{\pm\pm}$ could be probed in $e^- e^+$ collider from $Z^\prime$ using 4$\mu$ final states, however, for $x_H=0$ and 1 the final state involving 4$e$ signal in IH case could be observed with nearly 5$\sigma$ significance for $m_3=0.01$ eV and 0.001 eV respectively due to the large branching ratio of $\Delta^{\pm \pm}$ into 2$e$ mode $(46\%$ and $48\%)$. Such a final state will be prominently studied in NH case due to low branching ratio of $\Delta^{\pm \pm}$ into 2$e$ mode  $(3\%$ and $0.21\%)$. From Fig.~\ref{fig:Xsec-2-1} we find that $\Delta^{\pm \pm}$ production cross section from purely SM process will be nearly 0.005 fb. After the decay of $\Delta^{\pm \pm}$ and comparing with the SM background, the prediction of probing $\Delta^{\pm \pm}$ from purely SM processes will be could be below 2$\sigma$. As a result $Z^\prime$ mediation could have interesting role in probing $\Delta^{\pm \pm}$ and neutrino mass hierarchy associated with it.

The $\Delta^{\pm \pm}$ production cross sections at muon collider are given in the lower panel of Fig.~\ref{fig:Xsec-2-1}. 
Using Eq.~(\ref{evt}) we estimate signal events for $M_{Z^\prime}=10$ TeV and an integrated luminosity of 1 ab$^{-1}$ to study $4e$ and $4\mu$ final states 
in the NH and IH cases for $m_1=0.01$ eV and $0.001$ eV, taking $x_H=-1$, 0 and 1. 
In the NH case with $x_H=-1$, we estimate 540 events for $4e$ final sate signal taking $m_1=0.01$ eV 
whereas that for $m_1=0.001$ eV is negligibly small due to the small branching ratio of $\Delta^{\pm\pm}$. 
For the $4\mu$ final state for $m_1=0.01$ $(0.001)$ eV, we find $6.94\times 10^{4}$ $(5.4\times 10^{4})$ events. 
We also find that $2e2\mu$ final state will have nearly 540 events for $m_1=0.01$ eV and $0.001$ eV. 
For $\mu^- \mu^+$ collider, we generate the generic SM backgrounds for the process $\mu^- \mu^+ \to \ell^+ \ell^- \ell^+ \ell^-$ 
with $\ell= e,~\mu$ at $\sqrt{s}=10$ TeV and find the cross sections to be $10^{-6}$ pb, $10^{-6}$ pb and $10^{-4}$ pb for $4\mu$, $4e$ and $2e2\mu$ final states, respectively. 
As a result, the SM background events are only $\mathcal{O}(1-100)$ at 1 ab$^{-1}$ luminosity. 
Therefore, the signal events can be observed with more than $5\sigma$ significance for the NH case. 
For the IH case, we find $1.23\times 10^{5}$ $(1.4\times 10^{5})$ events for the $4e$ final state for $m_3=0.01$ $(0.001)$ eV 
whereas $1.734\times 10^{4} (9.08\times 10^{3})$ events for the $4\mu$ final state. 
For the $2e2\mu$ final state, we find 135 (86) events for $m_3=0.01$ (0.001) eV. 
Estimating the signal and SM backgrounds, we find that in case of muon collider these signals could be observed with more than $5\sigma$ significance. 
As mentioned above, our analysis did not consider the $\tau$ induced final states as we are not analyzing the decay of $\tau$ lepton into jets and lighter leptons. However, in future we may study such possibilities in detail. 
In case of $x_H=0$ and $1$, we find that $\Delta^{\pm \pm}$ production cross sections are nearly three times more than those for $x_H=-1$. 
In the same line, we can predict that for $x_H=0$ and $1$ finding $4e$, $4\mu$ and $2e2\mu$ signals for $M_{Z^\prime}=10$ TeV and $m_{\Delta^{\pm\pm}}=1.03$ TeV will be enhanced by nearly a factor of three to observe the signals with a significance more than $5\sigma$ at muon colliders in future. 
Besides the $Z^\prime$ induced production, we study the purely SM gauge boson mediated $\Delta^{\pm\pm}$ production and 
find that the corresponding cross section is about 3 orders of magnitude smaller than the once from the $Z^\prime$ induced process. 
As discussed above, the SM background events are 1-2 orders of magnitude smaller than the signal events. 
As a result, we predict that such final states could be probed in a muon collider with a significance of 5$\sigma$ depending on the neutrino mass hierarchy. 

\begin{figure}
\centering
\includegraphics[scale=0.4]{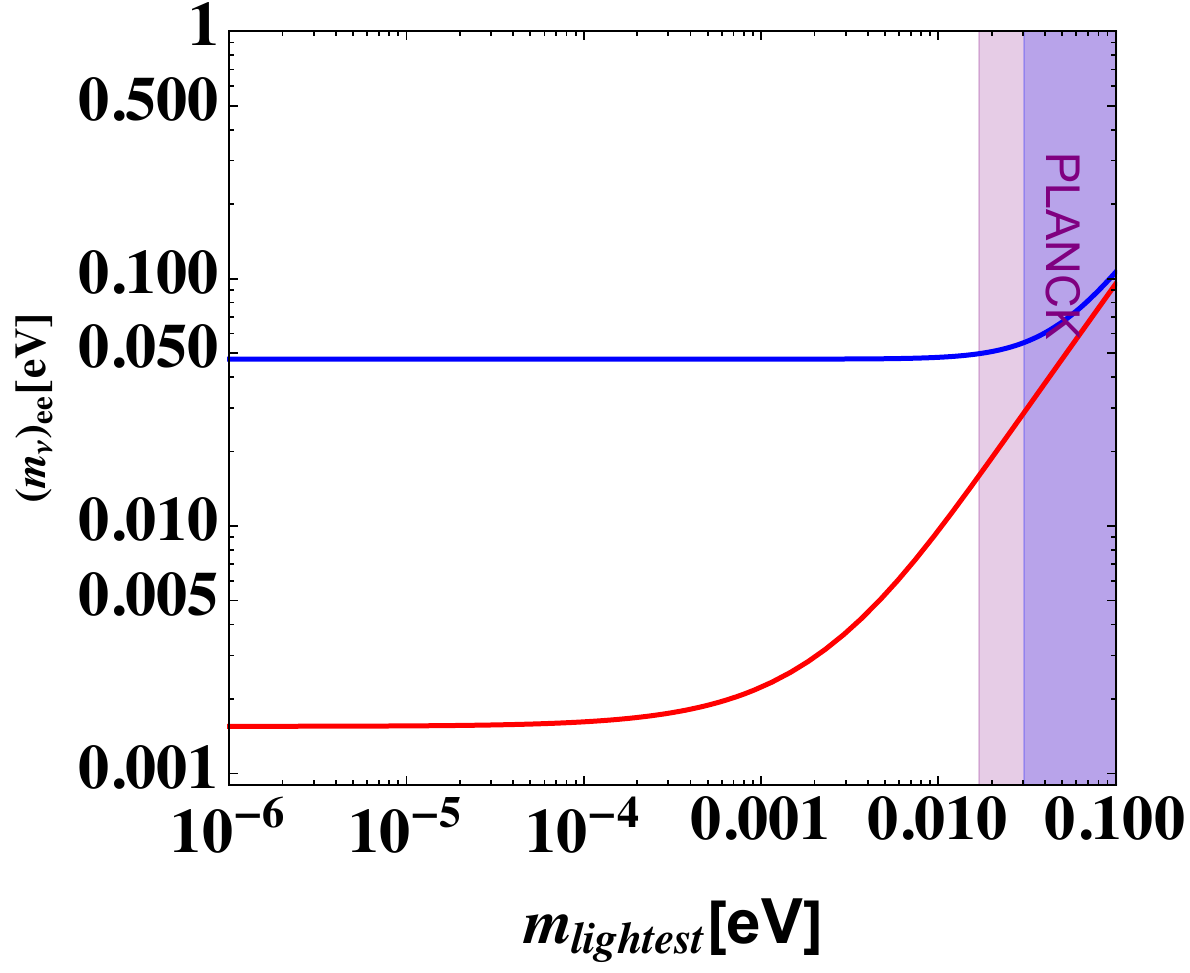}
\caption{Effective neutrino mass for the neutrinoless double beta decay in NH (red) and IH (blue) cases as a function of lightest light neutrino mass. Shaded regions are excluded by PLANCK data for NH (blue) and IH (purple) cases.} 
\label{fig:BRtrip3}
\end{figure}
In Fig.~\ref{fig:BRtrip3} we show the effective neutrino mass from the $(1,1)$ element of the neutrino mass matrix, responsible for the neutrinoless double beta decay, with respect to the lightest light neutrino mass eigenvalue for the NH and IH cases 
by red and blue solid lines, respectively, along with the PLANCK exclusion limits. 
We see that the effective mass shows its strong dependence on $m_1$ for the NH case, 
which is correlated to the strong dependence of BR($\Delta^{--} \to e^- e^-$) on $m_1$ 
for $m_1 \gtrsim 10^{-4}$ eV as shown in Fig.~\ref{fig:BRtrip1}.

\section{Conclusions} 
We study type-II seesaw scenario under an anomaly free, general $U(1)_X$ extension of the SM,
where Majorana-type light neutrino mass is generated by the VEV of the $SU(2)_L$ triplet scalar being charged under the $U(1)_X$ gauge group. 
In this scenario, after the breaking of the $U(1)_X$ gauge symmetry, the $U(1)_X$ gauge boson $Z^\prime$, which is electric-charge neutral and 
interacts differently with left- and right-handed SM fermions ($x_H \neq 0$), acquires its mass.   
Due to its $U(1)_X$ charge, the doubly-charged component in the triplet scalar can be produced at the high energy colliders 
such as LHC, FCC, $e^-e^+$ and $\mu^-\mu^+$ at 14 TeV, 100 TeV, a few TeV and ${\cal O}$(10 TeV) center of mass energies, respectively,
not only through the SM gauge bosons (photon and $Z$ boson) but also through a resonant production of  the $Z^\prime$ boson followed by its decay 
to a pair of doubly-charged scalars.   
In our scenario, the doubly-charged scalar dominantly decays to same-sign leptons and its branching ratios to lepton flavors are directly connected 
to the neutrino oscillation data, depending on the NH and IH neutrino mass spectrum.   
Therefore, once the doubly-charged scalar has been discovered at future collider experiments, we can probe the type-II seesaw mechanism 
by measuring the branching ratios of  the doubly-charged scalar to lepton flavors. 

In our collider study on the doubly-charged scalar, we have investigated the scalar production induced from a resonant $Z^\prime$ boson production 
followed by its decay to a pair of doubly-charged scalars. 
For 4 lepton final states from the decays of a pair of doubly-charged scalars, 
we have focused on $4e$, $4\mu$ and $2e2\mu$ signal events 
at LHC with 3 ab$^{-1}$ luminosity, FCC with 3 ab$^{-1}$ luminosity, and $e^-e^+$ and $\mu^- \mu^+$ colliders at 1 ab$^{-1}$ luminosity, respectively. 
Selecting the invariant mass of the final states to be in the vicinity of the $Z^\prime$ boson mass, we have calculated the signal events 
and compered them with the signal events from the SM gauge boson mediated processes as well as the generic SM backgrounds. 
We have found that the signal events induced by $Z^\prime$ boson production always dominate over those from the SM gauge boson mediated processes. 
In case of the LHC, we have found that $4\mu$ and $4e$ final states can be observed with more than $2\sigma$ significance, 
depending on the NH and IH cases. 
In case of FCC, we have found that the significance of finding $4e$, $4\mu$ final states can be more than $5\sigma$ for the NH and IH cases; 
the $2e2\mu$ final state can be observed with a significance of $5\sigma$ for the NH case, while the significance in the IH case is found to be $3-4\sigma$. 
For $e^- e^+$ and $\mu^-\mu+$ colliders, we have found that all $4e$, $4\mu$ and $2e2\mu$ signal events can be observed with a significance of $5\sigma$ for both NH and IH cases. 
 
\section*{acknowledgments}
The works of PD and NO are supported by the United States Department of Energy Grant No. DE-SC0012447.
\begin{widetext}
\appendix
\section{Scalar potential, partial decay widths of $Z^\prime$ and $\Delta^{- -/++}$}
The scalar potential of the model is given by
\bea
V&=&-m_H^2 |H|^2+m_{\tilde{H}}^2 |\tilde{H}|^2- m_\Phi^2 |\Phi|^2+ m_{\Delta}^2 Tr(\Delta^\dagger \Delta)+ \frac{1}{2} \lambda_H |H|^4+
 \frac{1}{2} \lambda_{\tilde{H}} |\tilde{H}|^4+\frac{1}{2} \lambda_\Phi |\Phi|^4+\frac{1}{2} \lambda_\Delta (Tr(\Delta^\dagger \Delta))^2 \nonumber \\
 &+&\frac{1}{2} \tilde{\lambda}_\Delta ((Tr(\Delta^\dagger \Delta))^2-Tr((\Delta^\dagger \Delta)^2))
 +\lambda_{H\tilde{H}} |H|^2 |\tilde{H}|^2+ \tilde{\lambda}_{H\tilde{H}} |\tilde{H}^\dagger H''|^2+\lambda_{\Phi H} |\Phi|^2 |H|^2+ \lambda_{\Phi \tilde{H}} |\Phi|^2 |\tilde{H}|^2 \nonumber \\
 &+& \lambda_{\Phi \Delta} |\Phi|^2 Tr(\Delta^\dagger \Delta)+ \{\lambda_{HH} |H|^2+ \lambda_{\tilde{H}\tilde{H}} |\tilde{H}|^2\} Tr(\Delta^\dagger \Delta)+\lambda_{\tilde{H}\Delta} \tilde{H}^\dagger[\Delta^\dagger, \Delta]\tilde{H}+\lambda_{H\Delta} H^\dagger[\Delta^\dagger, \Delta]H \nonumber \\
 &-&\frac{\mu_1}{\sqrt{2}} (\tilde{H}^T.\Delta \tilde{H}+H.c.)+ (\lambda \Phi(\tilde{H}^\dagger H)+ H.c.).
\label{eq.pot}
\eea
Substituting $\langle \Phi \rangle =v_\Phi/\sqrt{2}$, the effective scalar potential at energies below  $v_\Phi$ is described as
\bea
V&=&-\tilde{m}_H^2 |H|^2+\tilde{m}_{\tilde{H}}^2 |\tilde{H}|^2+ \tilde{m}_{\Delta}^2 Tr(\Delta^\dagger \Delta)+ \frac{1}{2} \lambda_H |H|^4+
 \frac{1}{2} \lambda_{\tilde{H}} |\tilde{H}|^4+\frac{1}{2} \lambda_\Delta (Tr(\Delta^\dagger \Delta))^2+\frac{1}{2} \tilde{\lambda}_\Delta ((Tr(\Delta^\dagger \Delta))^2 \nonumber \\
 &-&Tr((\Delta^\dagger \Delta)^2))
 +\lambda_{H\tilde{H}} |H|^2 |\tilde{H}|^2+ \tilde{\lambda}_{H\tilde{H}} |\tilde{H}^\dagger H|^2+ \{\lambda_{HH} |H|^2+ \lambda_{\tilde{H}\tilde{H}} |\tilde{H}|^2\} Tr(\Delta^\dagger \Delta)+\lambda_{\tilde{H}\Delta} \tilde{H}^\dagger[\Delta^\dagger, \Delta]\tilde{H} \nonumber \\ 
 &+&\lambda_{H\Delta} H^\dagger[\Delta^\dagger, \Delta]H- \frac{\mu_1}{\sqrt{2}} (\tilde{H}^T.\Delta \tilde{H}+H.c.)+ (\tilde{m}_{H\tilde{H}}^2(\tilde{H}^\dagger H)+ H.c.). 
\label{eq1.pot}
\eea
Stationary conditions from Eq.~(\ref{eq1.pot}) lead to
\bea
\tilde{m}_{H}^2&=&\frac{\lambda_H v_H^3 + v_{\tilde{H}}^2 v_H(\lambda_{H\tilde{H}}+\tilde{\lambda}_{H\tilde{H}})+ v_H v_\Delta^2(\lambda_{HH}-\lambda_{H\Delta})-2\tilde{m}_{H\tilde{H}}^2 v_{\tilde{H}} }{2v_{H}}, \nonumber \\
\tilde{m}_{\tilde{H}}^2&=&\frac{2 \tilde{m}_{H\tilde{H}}^2 v_H^2-v_{\tilde{H}}(\lambda_{\tilde{H}} v_{\tilde{H}}^2+ v_H^2(\lambda_{H\tilde{H}}+\tilde{\lambda}_{H\tilde{H}})+v_\Delta^2(\lambda_{\tilde{H}\tilde{H}}-\lambda_{\tilde{H}\Delta})-2\mu_1 v_\Delta )}{2v_{\tilde{H}}}, \nonumber \\
\tilde{m}_{\Delta}^2&=&\frac{\mu_{1} v_{\tilde{H}}^2- v_{\Delta} (\lambda_{H\tilde{H}}v_\Delta^2+ v_{\tilde{H}}^2 (\lambda_{\tilde{H}\tilde{H}}-\lambda_{\tilde{H} \Delta})+ v_H^2(\lambda_{HH}-\lambda_{H\Delta}) )}{2 v_\Delta}, 
\label{eq.st}
\eea
where $v_{H, \tilde{H}}$ are the VEVs of $H$ and $\tilde{H}$, respectively, satisfying $v=\sqrt{v_H^2+v_{\tilde{H}}^2}=246$ GeV. 
We can also write 
\bea
\mu_1=\frac{v_{\Delta}(\lambda_{\tilde{H}} v_{\Delta}^2+ v_{\tilde{H}}^2(\lambda_{\tilde{H}\tilde{H}}-\lambda_{\tilde{H} \Delta})+v_H^2(\lambda_{HH}-\lambda_{H\Delta})+2\tilde{m}_\Delta^2)}{v_{\tilde{H}}^2}.
\label{eq.type-II}
\eea
The type-II seesaw mechanism can be achieved from the approximation $\tilde{m}_{\Delta}^2 \gg v$ reducing to $v_\Delta\simeq \frac{\mu_1 v_{\tilde{H}}^2}{2 \tilde{m}_{\Delta}^2}$, and small $\mu_1$ and $v_{\tilde{H}}^2$ lead to a small $v_\Delta$. 
The masses of the scalars are given by
\bea
m_{h/\tilde{h}}^2&=& \begin{pmatrix} \lambda_{\tilde{H}} v_{\tilde{H}}^2+ m_{\tilde{H} H}^2 \frac{v_H}{v_{\tilde{H}}}& v_{\tilde{H} v_H} (\lambda_{H\tilde{H}}+\tilde{\lambda}_{H\tilde{H}})-m_{\tilde{H} H}^2 \\
v_{\tilde{H} v_H} (\lambda_{H\tilde{H}}+\tilde{\lambda}_{H\tilde{H}})-m_{\tilde{H} H}^2&\lambda_{H} v_{H}^2+ m_{\tilde{H} H}^2 \frac{v_{\tilde{H}}}{v_H}\end{pmatrix}, \nonumber \\
m_{H}^2&=&m_{H^\pm}^2-\frac{1}{2}(\lambda_{\tilde{H}\Delta} v_{\tilde{H}}^2+\lambda_{H\Delta} v_{H}^2)=m_{A}^2, \nonumber \\
m_{\tilde{A}}^2&=& m_{\tilde{H}H}^2 (\frac{v_{\tilde{H}}}{v_H}+\frac{v_H}{v_{\tilde{H}}}), \nonumber \\
m_{\tilde{H}^\pm}^2&=& m_{\tilde{A}}^2- \frac{\tilde{\lambda}_{H\tilde{H}} v^2}{2},\nonumber \\
m_{H^\pm}^2&=& \frac{1}{2} (\lambda_{\tilde{H}\tilde{H}} v_{\tilde{H}}^2+ \lambda_{HH} v_H^2+2 m_\Delta^2),\nonumber \\
m_{\Delta^{\pm\pm}}&=& m_{H^\pm}^2+\frac{1}{2}(\lambda_{\tilde{H}\Delta} v_{\tilde{H}}^2+ \lambda_{H\Delta} v_{H}^2). 
\label{eq.mass}
\eea

The partial decay widths of $Z^\prime$ and doubly charged component in the triplet scalar $(\Delta^{\pm \pm})$ are given by
\bea
\sum_{f= \rm quarks, leptons}\Gamma(Z^\prime \to f \overline{f})&=& \frac{g_X^2}{24 \pi} M_{Z^\prime} (13+16x_H+10x_H^2) \nonumber \\ 
\Gamma(Z^\prime \to \Delta^{\pm \pm} \Delta^{\mp \mp}/\Delta^{+} \Delta^{-})&=&
\frac{g_X^2 (x_H+2)^2 M_{Z^\prime}}{48 \pi} \Big[1-4\frac{m_{\Delta^{++}}^2}{M_{Z^\prime}^2}\Big]^{\frac{3}{2}}\nonumber \\
\Gamma(Z^\prime \to N_R \overline{N_R})&=& \frac{g_X^2}{24 \pi} M_{Z^\prime} x_{N_R}^2 \Big[1-4\frac{m_N^2}{M_{Z^\prime}}\Big]^{\frac{3}{2}},(x_{N_R}=-4, 5), 
\label{pdw}
\eea
\bea
\Gamma(\Delta^{\pm\pm} \to \ell_i^\pm \ell_j^\pm)&\simeq& C_{ij}\frac{|Y^{ij}|^2}{2}\frac{m_{\Delta^{\pm\pm}}}{4 \pi},  \nonumber \\
\Gamma(\Delta^{\pm \pm} \to W^\pm W^\pm) &\simeq& \frac{g_X^4 v_\Delta^2 m_{\Delta^{\pm \pm}}^2}{64 \pi m_W^4} 
\Big(3\frac{m_{W}^4}{m_{\Delta^{\pm \pm}}^4}-\frac{m_W^2}{m_{\Delta^{\pm \pm}}^2}+\frac{1}{4}\Big), 
\label{decayXX}
\eea
where $C_{ij}=1 (\frac{1}{2})$ for $i\neq j$ $(i=j)$.\\
\end{widetext}
\vspace{-0.398in}
\bibliographystyle{utphys}
\bibliography{reference}
\end{document}